\documentclass{elsart}

\usepackage[square,comma]{natbib}
\usepackage{graphicx}
\usepackage{pxfonts}

\usepackage{amssymb}

\begin{document}

\thispagestyle{empty}
\begin{Large}
\textbf{DEUTSCHES ELEKTRONEN-SYNCHROTRON}

\textbf{\large{in der HELMHOLTZ-GEMEINSCHAFT}\\}
\end{Large}

DESY 06-222

December 2006

\begin{eqnarray}
\nonumber &&\cr \nonumber && \cr \nonumber &&\cr
\end{eqnarray}
\begin{eqnarray}
\nonumber
\end{eqnarray}
\begin{center}
\begin{Large}
\textbf{Longitudinal Wake Field for an Electron Beam Accelerated
through a Ultra-High Field Gradient}
\end{Large}
\begin{eqnarray}
\nonumber &&\cr \nonumber && \cr
\end{eqnarray}

\begin{large}
Gianluca Geloni, Evgeni Saldin, Evgeni Schneidmiller and Mikhail
Yurkov
\end{large}
\textsl{\\Deutsches Elektronen-Synchrotron DESY, Hamburg}
\begin{eqnarray}
\nonumber
\end{eqnarray}
\begin{eqnarray}
\nonumber
\end{eqnarray}
\begin{eqnarray}
\nonumber
\end{eqnarray}
ISSN 0418-9833
\begin{eqnarray}
\nonumber
\end{eqnarray}
\begin{large}
\textbf{NOTKESTRASSE 85 - 22607 HAMBURG}
\end{large}
\end{center}
\clearpage
\newpage

\begin{frontmatter}



\title{Longitudinal Wake Field for an Electron Beam Accelerated through a Ultra-High Field Gradient}


\author[DESY]{Gianluca Geloni}
\author[DESY]{Evgeni Saldin}
\author[DESY]{Evgeni Schneidmiller}
\author[DESY]{Mikhail Yurkov}

\address[DESY]{Deutsches Elektronen-Synchrotron (DESY), Hamburg,
Germany}

\begin{abstract}
Electron accelerators with higher and higher longitudinal field
gradients are desirable, as they allow for the production of high
energy beams by means of compact and cheap setups. The new
laser-plasma acceleration technique appears to constitute the more
promising breakthrough in this direction, delivering unprecedent
field gradients up to TV/m. In this article we give a quantitative
description of the impact of longitudinal wake fields on the
electron beam. Our paper is based on the solution of Maxwell's
equations for the longitudinal field. Our conclusions are valid
when the acceleration distance is much smaller than the the
overtaking length, that is the length that electrons travel as a
light signal from the tail of the bunch overtakes the head of the
bunch. This condition is well verified for laser-plasma devices.
We calculate a closed expression for the impedance and the wake
function that may be evaluated numerically. It is shown that the
rate of energy loss in the bunch due to radiative interaction is
equal to the energy emitted through coherent radiation in the
far-zone. Furthermore, an expression is found for the asymptotic
limit of a large distance of the electron beam from the
accelerator compared with the overtaking length. Such expression
allows us to calculate analytical solutions for a Gaussian
transverse and longitudinal bunch shape. Finally, we study the
feasibility of Table-Top Free-Electron Lasers in the Vacuum
Ultra-Violet (TT-VUV FEL) and X-ray range (TT-XFEL), respectively
based on $100$ MeV and $1$ GeV laser-plasma accelerator drivers.
Numerical  estimations presented in this paper indicate that the
effects of the time-dependent energy change induced by the
longitudinal wake pose a serious threat to the operation of these
devices.
\end{abstract}


\end{frontmatter}


\clearpage

\section{\label{sec:intro} Introduction}

The quest for the ultimate electron accelerator will never end. A
better beam quality, a higher energy, a smaller size and lower
costs will always be goals to be pursued. In particular, size and
costs are strictly related. This justifies research activities
towards higher and higher longitudinal field gradients. The
availability of higher longitudinal field gradients is obviously
related, in its turn, to the possibility of production of beams
with higher energies.

At the present day, particle accelerator technology is mainly
based on Radio-Frequency (RF) devices. However, in the past few
years a novel laser-plasma acceleration method has been developed
and experimentally demonstrated \cite{MANG, GEDD, FAUR}, that
promises to outclass all existing accelerator technologies. The
laser-plasma acceleration technique requires an ultrashort (a few
femtoseconds long) high intensity (Tera to Peta-Watt) laser pulse
focused into a supersonic Helium gas jet (or filled capillary)
with density around $10^{19}/cm^3 $. This produces a region free
of electrons propagating behind the laser pulse. Many electrons
(up to $10^{10}$) are captured in this electron-free zone after
about one plasma oscillation, and eventually accelerated by the
huge electric field produced by the positive ion background with
gradients up to TV/m. Up-to-date experimental verifications
demonstrated gradients of the order of $100$ GeV/m, that can
potentially accelerate electrons up to energies in the GeV-range
within a few millimeters.

In this paper we study the important issue of longitudinal wake
fields produced within electron beams accelerated with
high-gradient fields. We assume that the acceleration distance
$d_a$ is much smaller than the the overtaking length. This is the
distance travelled by the electrons as a light signal from the
tail of the bunch overtakes the head of the bunch. Given a bunch
of rms length $\sigma_z$, the overtaking length can thus be
written as $2 \gamma^2 \sigma_z$, and corresponds to the radiation
formation length $2 \gamma^2 \lambdabar$ calculated at $\lambdabar
= \sigma_z$, $\lambdabar = \lambda/(2\pi)$ being the reduced
radiation wavelength. When $d_a \ll 2\gamma^2 \sigma_z$, the
electrons can be assumed to be accelerated at a single position
$z_A$ down the beamline. This is the case for laser-plasma
devices, since acceleration in the GeV range takes place within a
few millimeters only. However, it is not the case for conventional
accelerators, that feature typical gradients up to a few tens of
MeV/m, and thus need several tens of meters to reach the GeV
range. The assumption $d_a \ll 2\gamma^2 \sigma_z$ greatly
simplifies wake calculations. In particular, when this condition
is verified, the wake generated along the part of the trajectory
following the acceleration point $z_A$ is independent of any
detail of the particular realization of the accelerator. In this
sense, our study is fundamental, because it remains valid
independently of the particle accelerator technology chosen,
provided that $d_a \ll 2\gamma^2 \sigma_z$. One may also have
contributions to the wake generated along the part of the
trajectory following the acceleration point $z_A$. These
contributions depend on the physical nature of the accelerator,
can be separately calculated, and will be neglected in this paper
because they do not affect the bunch in the case of a laser-plasma
accelerator.

We base our study on the solution of Maxwell's equations for the
longitudinal field. With the help of the fundamental example of an
electron travelling in uniform motion we show that the paraxial
approximation can be applied to describe electromagnetic sources
up to the observation point, while sources after the observation
point in the beam propagation direction can be neglected. Then we
make use of the paraxial approximation to calculate the Fourier
transform of the longitudinal field produced by fixed
electromagnetic sources (current and charge densities) at a
certain observation plane down the beamline. This expression is at
the basis of our treatment, because it allows to calculate a
closed expression for the longitudinal impedance and for the wake
function to be evaluated numerically. The knowledge of the real
part of the impedance further allows verification of the energy
conservation principle. Impedance and wake function yield an
analytical expression in the asymptotic limit of a large distance
of the electron beam from the accelerator compared with the
overtaking length, i.e. in the far-field zone for all wavelengths
of interest (up to $\lambdabar \sim \sigma_z$). This asymptotic
limit is of practical relevance, as it allows simple estimations
of the impact of the longitudinal wake on the electron beam energy
change.

We give an application of our theoretical work by studying the
feasibility of Free-electron Lasers (FELs) drivers, proposed as a
possible use of laser-plasma accelerators.

Coherent sources of electromagnetic radiation have become very
important research tools in science and industry. Recent advance
in conventional (RF) particle accelerator techniques allow to
construct and operate Free-Electron Lasers in the VUV as well as
in the X-ray range. Lasing at wavelengths shorter than the
ultra-violet can be achieved with a single-pass, high-gain FEL
amplifier. Because of the lack of powerful, coherent seeding
sources, short-wavelength FELs work in the so-called
Self-Amplified Spontaneous Emission (SASE) mode, where the
amplification process starts from shot noise in the electron beam
\cite{SAS1,SAS2,SAS3}. Experimental realization of SASE FELs
developed very rapidly during the last decade. The shortest
wavelength ever generated by an FEL, $\lambda = 13$ nm, has been
achieved in 2006 at FLASH (Free-electron LAser in Hamburg).
Regular user operation of FLASH started in 2005 \cite{FLA1}.
Currently, this facility produces GW-level laser-like radiation
pulses with $10$ to $50$ fs duration in the wavelength range from
$13$ to $45$ nm. Recently, the German government, encouraged by
these results, approved funding a hard X-ray SASE FEL (XFEL) user
facility, the European XFEL \cite{XFEL}. The US department of
Energy (DOE) has approved the start of construction of the Linac
Coherent Light Source (LCLS) at the Stanford Linear Accelerator
Center (SLAC) \cite{SLAC}. The LCLS and the European XFEL project
are scheduled to start operation in 2009 and 2013 respectively.

With the recent progress in the development of laser-plasma
accelerators \cite{MANG, GEDD, FAUR} a discussion started about
wether and how this technology can be used to provide a
cost-effective driver for SASE FELs. The very short length scale
of plasma acceleration may eventually allow for the development of
very compact laser-like sources both in the VUV and in the X-ray
range. In this paper we will not address the delicate issue of
technical realization of such table-top FEL drivers. These should
finally provide short bunches up to about $10$ fs duration in the
$100$ kA current class (i.e. with a charge of about $1$ nC),
featuring a relative energy spread of $0.1\%$ for beam energies in
the GeV range and a normalized emittance in the order of $1$
mm$\cdot$mrad. Instead, we will restrict our discussion to
fundamental electrodynamical questions, supposing that
laser-plasma drivers already exist. Once this assumption is made,
there are two ways of taking advantage of such device.

The first way is conservative. Up-to-date laser-plasma technology
can produce beam energies up to about $1$ GeV. With this energy,
one may use parameters for the system electron beam-undulator in
the same range of those at FLASH. In this fashion one may
eventually achieve wavelengths up to $5 \div 10$ nm. This
application amounts to substitution of the conventional
accelerator system at FLASH with the table-top laser-plasma
driver. This would not lead to a table-top VUV-FEL, but it would
certainly reduce its size. The only condition for this scheme to
work is that a laser-plasma driver capable of providing the same
beam parameters as the accelerator system at FLASH must first be
built in reality.

The second way is more ambitious, and consists in using
laser-plasma accelerators in the GeV range to build Table-Top
X-ray FELs by using a completely different set of parameters for
the system electron beam-undulator. Namely, an order of magnitude
smaller electron energy, an order of magnitude larger peak
current, an order of magnitude smaller undulator parameter and an
order of magnitude shorter undulator period. The present study
allows to estimate the influence of longitudinal wake fields on
the relative energy change of the electron beam. In particular, we
estimate the magnitude of the induced correlated energy-change as
a function of the position within the bunch, as electrons travel
through the undulator. This is a fundamental effect that cannot be
avoided by fine tuning of the parameters of the setup. Moreover,
the energy change grows during the passage of the beam through the
undulator, so that one cannot assume a fixed correlated energy
change through the undulator. Our conclusion is that the magnitude
of this effect poses a serious threat to the feasibility of a
table-top FEL in the x-ray range (TT-XFEL), as well as in the VUV
range (TT-VUV FEL).

Our work is organized as follows. First, in Section
\ref{sec:lfield} we calculate the longitudinal field in the
space-frequency domain. In the following Section \ref{sec:wakpoi}
we introduce the basic quantities impedance and wake function, and
we review their relation with the Poynting theorem. In Section
\ref{sec:asymp} we present an analytical expression for the
asymptotes of the longitudinal impedance in the limit for a large
longitudinal distance of the beam with respect to the overtaking
length. We further verify the energy conservation principle in
Section \ref{sec:conse}. Section \ref{sec:enchange} is dedicated
to the calculation of an analytical expression for the asymptotes
of the longitudinal wake, always in the limit for a large
longitudinal distance of the beam with respect to the overtaking
length. Finally, our study about the feasibility of the TT-VUV FEL
and the TT-XFEL is given in Section \ref{sec:example}. Section
\ref{sec:conc} concludes our work with a few final remarks.

\section{\label{sec:lfield} Calculation of the longitudinal field
in the space-frequency domain}

\begin{figure}
\begin{center}
\includegraphics*[width=140mm]{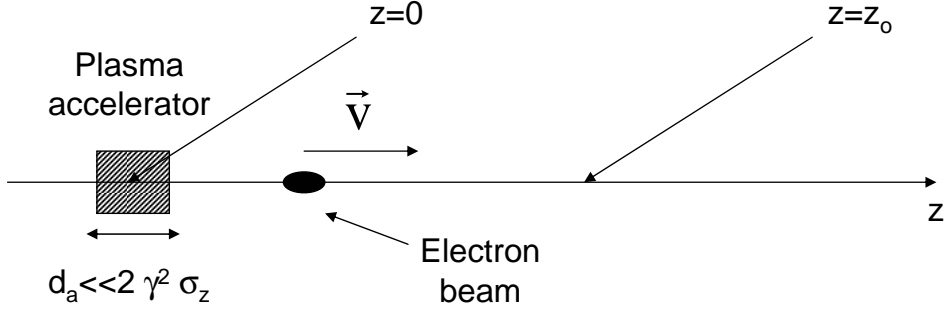}
\caption{\label{geom1} Geometry.}
\end{center}
\end{figure}
Consider the scheme in Fig. \ref{geom1}, representing the object
of our study. Consider position $z_A$, representing the exit of
the plasma accelerator. At this position, the nominal Lorentz
factor of electrons is $\gamma$. The accelerator switches on both
harmonics of the sources and of the field. Since we consider the
wake generated along the part of the trajectory following the
acceleration point $z_A$, the nature of the switcher is not
important. It is important, though, that the switching process
happens on a longitudinal scale $d_a \ll 2 \gamma^2 \lambdabar$,
representing the distance that a light signal has to travel before
it overtakes an electron of a distance $\lambdabar=c/\omega$, $c$
being the speed of light in vacuum and $\omega$ the angular
frequency of the radiation in the space-frequency domain.
Significant emission of radiation in vacuum is present for
wavelengths longer than the longitudinal bunch length $\sigma_z$,
up to transverse beam sizes $\sigma_\bot \lesssim \gamma \sigma_z$
as we will see. As a result, for typical ultra-relativistic beams
the condition $d_a \ll 2 \gamma^2 \lambdabar$ can be read $d_a \ll
2 \gamma^2 \sigma_z$, $2\gamma^2 \sigma_z$ being the overtaking
length.

We begin our investigation with the help of general knowledge of
electromagnetic theory \cite{JACK}, by calculating the
longitudinal electric field in the space-frequency
domain\footnote{In this paper we define Fourier transform and
inverse transform pair of a function $f(t)$ as

\begin{eqnarray}
\nonumber \bar{f}(\omega) = \int d t ~f(t) \exp\left[i\omega t
\right]~;~~f(t) = \frac{1}{2\pi} \int d \omega ~\bar{f}(\omega)
\exp\left[-i\omega t \right]~. \label{prip}
\end{eqnarray}
For future use, we also write, explicitly, the definitions of the
two-dimensional Fourier transform and inverse transform of a
function $g(\vec{r})$ in agreement with the one-dimensional
notation:

\begin{eqnarray}
\nonumber \tilde{g}(\vec{k}) = \int \d \vec{r} ~g(\vec{r})
\exp\left[i \vec{r} \cdot \vec{k} \right]~;~~{g}(\vec{r}) =
\frac{1}{4\pi^2} \int \d \vec{k} ~\tilde{g}(\vec{k}) \exp\left[-i
\vec{r} \cdot \vec{k} \right] , \label{spatgft}
\end{eqnarray}
the integration being understood over the entire plane. If $g$ is
circular symmetric we can introduce the Fourier-Bessel transform
and inverse transform pair:

\begin{eqnarray}
\nonumber \tilde{g}({k}) = 2\pi \int_0^{\infty} d{r}~ r g(r) J_o(k
r)~;~~ {g}({r}) = \frac{1}{2\pi} \int_0^{\infty} d{k}~ k
\tilde{g}(k) J_o(k r)~,\label{bessspatgft}
\end{eqnarray}
$r$ and $k$ indicating the modulus of the vectors $\vec{r}$ and
$\vec{k}$ respectively, and $J_o$ being the zero-th order Bessel
function of the first kind.} $\bar{E}_z(z_o, \vec{r}_{\bot o},
\omega)$ produced at frequency $\omega$ by given electromagnetic
sources in vacuum and detected at longitudinal position $z_o>z_A$
and transverse position $\vec{r}_o$ (see Fig. \ref{geom1}).

The field in the space-frequency domain $\vec{\bar{E}}(z_o,
\vec{r}_{\bot o}, \omega)$ is the Fourier transform of the field
in the space-time domain, $\vec{E}(z_o,\vec{r}_{\bot o},t)$. As is
well-known, for a fixed frequency $\omega=\omega_o$ the field in
the space-frequency domain is strictly related to the amplitude
$\vec{\bar{E}}_{\omega_o}(z_o,\vec{r}_{\bot o})$ of a
monochromatic field oscillating in time with angular frequency
$\omega$. In fact

\begin{eqnarray}
\vec{E}(z_o,\vec{r}_{\bot o},t) = \vec{\bar{E}}_{\omega_o}
(z_o,\vec{r}_{\bot o}) \exp\left[-i\omega_o t\right] + C.C.~,
\label{wavemo}
\end{eqnarray}
where "C.C." indicates the complex conjugate of the preceding
term. It follows that

\begin{eqnarray}
\vec{\bar{E}}(z_o, \vec{r}_{\bot o}, \omega_o) = 2 \pi
\vec{\bar{E}}_\omega (z_o,\vec{r}_{\bot o})
\delta(\omega-\omega_o)~.\label{rel0}
\end{eqnarray}
Therefore, one may use the field in the space-frequency domain for
a given frequency and yet think of the amplitude of a
monochromatic field or viceversa. In fact, although they have
different dimensions, the knowledge of one quantity fully
identifies the other through Eq. (\ref{rel0}). As an example,
$\vec{\bar {E}}$ satisfies the Helmholtz equation

\begin{equation}
c^2 \nabla^2 \vec{\bar{E}} + \omega^2 \vec{\bar{E}} = 4 \pi c^2
\vec{\nabla} \bar{\rho} - 4 \pi i \omega \vec{\bar{j}}~,
\label{trdisoo}
\end{equation}
where $\bar{\rho}(\vec{r},\omega)$ and
$\vec{\bar{j}}(\vec{r},\omega)$ are the Fourier transforms of the
space-time domain charge density $\rho(\vec{r},t)$ and current
density $\vec{j}(\vec{r},t)$. In the same way, an equation for
$\vec{\bar{E}}_\omega$ is found substituting $\vec{\bar{j}}$ and
$\rho$ with $\vec{\bar{j}_\omega}$ and $\rho_\omega$, related to
$\vec{\bar{j}}$ and $\bar{\rho}$ through an analogous of Eq.
(\ref{rel0}).

Eq. (\ref{trdisoo}) can be solved with the help of an appropriate
Green's function $G(z_o-z',\vec{r}_{\bot o}-\vec{r'}_\bot)$
yielding

\begin{eqnarray}
{\bar{E}}_z(z_o, \vec{r}_{\bot o},\omega) &=& - 4\pi
\int_{-\infty}^{\infty} dz' \int d \vec{r'}_{\bot}
\left(\frac{i\omega}{c^2} \bar{j}_z -\frac{\partial
\bar{\rho}}{\partial z'} \right)  G(z_o-z',\vec{r}_{\bot
o}-\vec{r'}_\bot) ~, \label{sol1}
\end{eqnarray}
the integration in $d \vec{r'}_{\bot}$ being performed over the
entire transverse plane. An explicit expression for the Green's
function to be used in Eq. (\ref{sol1}) is given by

\begin{eqnarray}
G(z_o-z',\vec{r}_{\bot o}-\vec{r'}_\bot) &=& -\frac{\exp\left\{ i
(\omega/c) \left[\left|\vec{r}_{\bot
o}-\vec{r'}_\bot\right|^2+\left(z_o-z'\right)^2\right]^{1/2}\right\}}{4\pi
\left[\left|\vec{r}_{\bot
o}-\vec{r'}_\bot\right|^2+\left(z_o-z'\right)^2\right]^{1/2}}
\label{greenhyp}~,
\end{eqnarray}
that automatically includes the proper boundary conditions at
infinity.

Since electrons are moving along the z-axis, we write the harmonic
components of the charge and current density, $\bar{\rho}$ and
$\bar{j}_z$ as

\begin{eqnarray}
\bar{\rho}\left(z', \vec{r'}_\bot, \omega \right) =
\rho_o\left(\vec{r'}_\bot \right) \bar{f}(\omega)
\exp\left[i\frac{\omega z'}{v}\right] u(z'-z_A)\label{composrho}
\end{eqnarray}
and

\begin{eqnarray}
\bar{j}_z\left(z', \vec{r'}_\bot, \omega \right) = \beta c
\bar{\rho}~. \label{composj}
\end{eqnarray}
Notation $u(z'-z_A)$ in Eq. (\ref{composrho}) indicates a
Heaviside step function centered at position $z_A$, whose presence
signifies that there are no sources before the plasma accelerator,
i.e. before the point $z'=z_A$. This describes a switch-on
process. When the switching distance $d \ll 2 \gamma^2 \sigma_z$,
the nature of the switcher is not important for the description of
the wake associated with Eq. (\ref{composrho}) and Eq.
(\ref{composj}), i.e. the wake generated along the part of the
trajectory following the acceleration point $z_A$. However, Eq.
(\ref{composrho}) and Eq. (\ref{composj}) alone violate the
continuity equation, and should be completed by
extra-contributions that depend on the nature of the switcher. For
example, if one thinks of an acceleration process where a low
energy bunch with Lorentz factor $\gamma_o$ is accelerated on a
distance $d$ up to a Lorentz factor $\gamma$, one should add to
Eq. (\ref{composrho}) and Eq. (\ref{composj}) the contribution of
the harmonic at frequency $\omega$ associated with the beam with
Lorentz factor $\gamma_o$. Such harmonic is characterized by a
different longitudinal velocity $\beta_o c$ with respect to $\beta
c$, and by a different wave number $k_o = \omega/v_o$. However,
its amplitude is the same of that after the acceleration process:

\begin{eqnarray}
\bar{\rho}_o\left(z', \vec{r'}_\bot, \omega \right) =
\rho_o\left(\vec{r'}_\bot \right) \bar{f}(\omega)
\exp\left[i\frac{\omega z'}{v_o}\right]
[1-u(z'-z_A)]\label{composrhoo}
\end{eqnarray}
and

\begin{eqnarray}
\bar{j}_{oz}\left(z', \vec{r'}_\bot, \omega \right) = \beta_o c
\bar{\rho}~. \label{composjoo}
\end{eqnarray}
As a result, the continuity equation is satisfied. However, Eq.
(\ref{composrhoo}) and Eq. (\ref{composjoo}) depend on the
particular process described (acceleration from $v_o$ to $v$), so
that their contribution depends on the particular switching
process selected. In this paper we will not consider this
contribution. We will focus, instead, on the switch-independent
part of the problem. Note that for the case of a plasma
accelerator, Eq. (\ref{composrhoo}) and Eq. (\ref{composjoo})
should be replaced by contributions due to a positive ion current
propagating in the opposite direction with respect to the
electrons. As a result we can neglect such source, that has no
effect on the longitudinal wake field acting on the electron beam.

The assumption of separability of variables $z'$ and
$\vec{r'}_\bot$, in Eq. (\ref{composrho}), together with the fact
that the function $\bar{f}$ is independent of $z'$ may be
satisfied as long as the transverse electron beam size
$\sigma_\bot$ remains unvaried along a formation length. This
happens for an angular divergence $\sigma_\theta$ such that
$\sigma_\bot \gg \sigma_\theta \gamma^2 \sigma_z$. Moreover, the
assumption of separability also requires that we can neglect
dynamical effects of self-interactions, because there is no
external force acting on the bunch. In this sense, we are
developing the zero-th order treatment of a perturbation theory,
where the perturbation to the particles dynamics is given by the
self-interaction within the electron bunch. The quantity $\rho_o$
has the meaning of transverse electron beam distribution; in
principle, it may depend on the harmonic $\omega$, but in the
following we will assume it does not. Thus, all information about
the longitudinal electron density distribution $f(t)$ is included
in the function $\bar{f}(\omega)$, that is its Fourier transform.
A typical Gaussian beam model, that will be useful later on, is
defined by

\begin{eqnarray}
\rho_o\left(\vec{r'}_\bot \right) = \frac{1}{2\pi \sigma_\bot^2 c}
\exp\left[-\frac{r^{'2}_\bot}{2 \sigma_\bot^2}\right]~,
\label{rhor}
\end{eqnarray}
$\sigma_\bot$ being the $rms$ beam transverse dimension and by

\begin{eqnarray}
f(t) = \frac{(-e)N}{\sqrt{2\pi} \sigma_t} \exp\left[-\frac{t^2}{2
\sigma_t^2}\right]~~~\longleftrightarrow~~~\bar{f}(\omega) = (-e)N
\exp\left[-\frac{\omega^2 \sigma_t^2}{2}\right]~, \label{ft}
\end{eqnarray}
where $N$ is the number of electrons in the beam and $(-e)$ the
electron beam charge. Moreover, $\sigma_t$ is the $rms$ bunch
duration, connected with the $rms$ bunch length by $\sigma_z =
\beta c \sigma_t$, so that in terms of lengths

\begin{eqnarray}
f(s) = \frac{(-e) N}{\sqrt{2\pi} \sigma_z} \exp\left[-\frac{s^2}{2
\sigma_z^2}\right]~. \label{fs}
\end{eqnarray}
We now account for Eq. (\ref{composrho}) and Eq. (\ref{composj})
and calculate the derivative of $\bar{\rho}$ with respect to $z'$
in Eq. (\ref{sol1}), but keep the implicit form for  $G$. We
obtain

\begin{eqnarray}
{\bar{E}}_z &=& \frac{4 \pi i\omega \bar{f}(\omega) }{\gamma^2 c}
\int d \vec{r'}_{\bot} \rho_o\left(\vec{r'}_\bot \right)
\int_{z_A}^{\infty} dz' \exp\left[i\frac{\omega z'}{v}\right]
G(z_o-z',\vec{r}_{\bot o}-\vec{r'}_\bot) \cr && + 4\pi
\bar{f}(\omega) \exp\left[i\frac{\omega z_A}{v}\right] \int d
\vec{r'}_{\bot}  \rho_o\left(\vec{r'}_\bot \right)
G(z_o-z_A,\vec{r}_{\bot o}-\vec{r'}_\bot)~. \label{sol2}
\end{eqnarray}
The integration domain in $dz'$ in Eq. (\ref{sol2}), that is
$[z_A,\infty)$, can be represented as $[z_A, z_o]\cup [z_o,
\infty)$. The integral in $dz'$ can be written as the sum of two
integrals in $dz'$ performed over the separate domains $[z_A,
z_o]$ and $[z_o, \infty)$. At this point we perform the following
operations. First, we neglect the integral over the interval
$[z_o, \infty)$, i.e. we neglect the effects of electromagnetic
sources located in $[z_o,\infty)$. Second, we apply the paraxial
approximation. This means that we solve, for sources located in
$[z_A, z_o]$, the paraxial equation

\begin{equation}
c^2 \exp\left[i\frac{\omega}{c} z\right] \left( \nabla_\bot^2
+\frac{2 i \omega}{c}{\partial\over{\partial z}}\right)
\vec{\widetilde{E}} = 4 \pi c^2 \vec{\nabla} \bar{\rho} - 4 \pi i
\omega \vec{\bar{j}}~, \label{parax}
\end{equation}
where we introduced the envelope of the field components

\begin{equation}
\vec{\widetilde{E}} = \vec{\bar{E}} \exp{\left[-i\omega
z/c\right]}~, \label{vtilde}
\end{equation}
because paraxial approximation implies a slowly varying envelope
of the field with respect to the wavelength $\lambda = 2\pi
c/\omega$. Note that the source term of Eq. (\ref{parax}) is now
multiplied by a phase factor $\exp[-i \omega z/c]$ with respect to
that of Helmholtz equation, Eq. (\ref{trdisoo}). As a result, we
can obtain an expression for ${\widetilde{E}}_z$ by formally
operating in Eq. (\ref{sol2}) in the following way. First, we
restrict the integration limits in $dz'$ to $[z_A, z_o]$. Second,
we substitute the Green's function $G$ with $\exp[-i \omega
z/c]\cdot G_p$, where $G_p$ is the Green's function for the
paraxial equation:

\begin{eqnarray}
G_p(z_o-z',\vec{r}_{\bot o}-\vec{r'}_\bot) = -\frac{1}{4\pi
(z_o-z')} \exp\left[ i\omega{\mid \vec{r}_{\bot o}
-\vec{r'}_\bot\mid^2\over{2c (z_o-z')}}\right] \label{green}~.
\end{eqnarray}
We then obtain the following expression for ${\widetilde{E}}_z$:

\begin{eqnarray}
&& {\widetilde{E}}_z(z_o, \vec{r}_{\bot o}, \omega) =\cr &&
-\frac{ i\omega \bar{f}(\omega) }{\gamma^2 c}  \int d
\vec{r'}_{\bot} \rho_o\left(\vec{r'}_\bot \right) \int_{z_A}^{z_o}
\frac{dz'}{(z_o-z')} \exp\left[\frac{i\omega z'}{2 \gamma^2 c} +
i\omega{\mid \vec{r}_{\bot o} -\vec{r'}_\bot\mid^2\over{2c
(z_o-z')}}\right] \cr && - \frac{\bar{f}(\omega)}{(z_o-z_A)}
\exp\left[\frac{i\omega z_A}{2 \gamma^2 c}\right]\int d
\vec{r'}_{\bot} \rho_o\left(\vec{r'}_\bot \right) \exp\left[
i\omega{\mid \vec{r}_{\bot o} -\vec{r'}_\bot\mid^2\over{2c
(z_o-z_A)}}\right] ~, \label{solgen}
\end{eqnarray}
We do not present here a general proof of the validity of Eq.
(\ref{solgen}). We rather verify its correctness in the particular
case $z_A \longrightarrow -\infty$, $z_o = 0$,
$\rho_o\left(\vec{r'}_\bot \right) = \delta \left(\vec{r'}_\bot
\right)$ and $\bar{f}(\omega) = (-e)$. This corresponds to Fourier
transform of the field of a particle in uniform motion, calculated
at position $(0,\vec{r}_{\bot o})$. In the chosen limit, Eq.
(\ref{solgen}) becomes

\begin{eqnarray}
&& {\widetilde{E}}_z(0, {r}_{\bot o}, \omega) = \frac{ i\omega
(-e)}{\gamma^2 c}   \int_{-\infty}^{0} dz' \frac{1}{z'} \exp\left[
\frac{i\omega z'}{2\gamma^2 c}-i\omega{{r}_{\bot o}^2\over{2c
z'}}\right] ~, \cr && \label{sol2bisprova0}
\end{eqnarray}
depending on the modulus of $\vec{r}_{\bot o}$ only. The integral
in Eq. (\ref{sol2bisprova0}) can be performed analytically using
$- \omega z'/(2\gamma^2 c)$ as integration variable in place of
$z'$ and taking advantage of the following relation, that is valid
for values $\alpha>0$:

\begin{eqnarray}
\int_{0}^{\infty} \frac{d x}{x}
\exp\left[i\left(-x+\frac{\alpha}{x}\right)\right] = 2
\int_{0}^{\infty} d\xi \frac{\cos\left(2 \sqrt{\alpha} \xi
\right)}{\sqrt{1+\xi^2}} = 2 K_o\left(2
\sqrt{\alpha}\right)~,\label{rele}
\end{eqnarray}
where $K_n$ indicates the n-th order modified Bessel function of
the second kind. We thus obtain the following expression:

\begin{eqnarray}
&& {\widetilde{E}}_z(0, {r}_{\bot o}, \omega) = - \frac{2 i\omega
(-e)}{\gamma^2 c} K_o\left(\frac{ \omega {r}_{\bot o}}{c \gamma}
\right) ~, \cr && \label{sol2bisprova}
\end{eqnarray}
in perfect agreement (aside for notational differences) with Eq.
(13.80) of \cite{JACK}, following directly from the Fourier
transform of the time-domain electric field.

To conclude this Section, we give an explicit presentation of Eq.
(\ref{solgen}) in the case $z_A \longrightarrow -\infty$, that
corresponds to the steady state solution for the electric field,
and for $z_A=0$, that represents the switch-on case without loss
of generality.

In the case $z_A \longrightarrow -\infty$ we can make still take
advantage of Eq. (\ref{rele}). Then, in the steady state limit,
Eq. (\ref{solgen}) becomes

\begin{eqnarray}
{\widetilde{E}}_z && (z_o, \vec{r}_{\bot o}, \omega) = - \frac{2
i\omega}{\gamma^2 c}  \exp\left[\frac{i\omega
z_o}{2\gamma^2c}\right] \bar{f}(\omega) \int d \vec{r'}_{\bot}~
\rho_o(\vec{r'}_\bot) K_o\left(\frac{\omega \mid \vec{r}_{\bot o}
-\vec{r'}_\bot\mid}{\gamma c}\right) ~.\cr && \label{solstdy}
\end{eqnarray}
It should be noted that in this case, the dependence of
$\widetilde{E}_z$ on $z_o$ is restricted to a phase factor only
(steady state solution). In the case $z_A = 0$, direct
substitution in Eq. (\ref{solgen}) yields

\begin{eqnarray}
&& {\widetilde{E}}_z(z_o, \vec{r}_{\bot o}, \omega) = -\frac{
\omega \bar{f}(\omega) }{\gamma^2 c} \int d \vec{r'}_{\bot}
\rho_o\left(\vec{r'}_\bot \right)\cr &&\times  \left\{ i
\int_{0}^{z_o} \frac{dz'}{(z_o-z')} \exp\left[\frac{i\omega z'}{2
\gamma^2 c} + i\omega{\mid \vec{r}_{\bot o}
-\vec{r'}_\bot\mid^2\over{2c (z_o-z')}}\right] +\frac{\gamma^2
c}{\omega z_o} \exp\left[ i\omega{\mid \vec{r}_{\bot o}
-\vec{r'}_\bot\mid^2\over{2c z_o}}\right]\right\} ~, \cr &&
\label{solzo}
\end{eqnarray}
It can be verified by inspection that, in the limit for $z_o \gg 2
\gamma^2 \lambdabar$, Eq. (\ref{solzo}) gives back the steady
state solution, Eq. (\ref{solstdy}). Eq. (\ref{solstdy}) and Eq.
(\ref{solzo}) or, more in general, Eq. (\ref{solgen}) can now be
used to estimate the impact of the field generated by the electron
beam on any particle in the beam. We do so with the help of the
concepts of wake fields and impedances, that are related to the
Poynting theorem as reviewed in the following Section.

\section{\label{sec:wakpoi}  Poynting theorem in the frequency
domain, longitudinal wake fields and impedances}

In the time domain, the Poynting theorem \cite{JACK} reads, at any
time $t$:

\begin{eqnarray}
\int_V \vec{{j}} \cdot \vec{{E}} ~d V+ \int_A \vec{{S}}\cdot
\hat{n}~ d A + \frac{d}{d t} \int_V \left({w}_e-{w}_m\right) ~d V
=0~. \label{PoyT}
\end{eqnarray}
Notations $V$ and $A$ in Eq. (\ref{PoyT}) respectively indicate
any finite volume and its surrounding surface, while $\hat{n}$
denotes the field of unit vectors normal to A, directed outwards.
In this way, the second integral in Eq. (\ref{PoyT}) represents
the flux of the Poynting vector, while the first and the second
integrals are volume integrals calculated within $V$. In Eq.
(\ref{PoyT}) we introduced the Poynting vector $\vec{S}$ and the
energy densities $w_e$ and $w_m$, understanding that they all are
functions of time, in the following way:

\begin{eqnarray}
\vec{{S}} = \frac{c}{4\pi} \left(\vec{{E}} \times
\vec{{B}}\right)~,~~~{w}_e = \frac{1}{8\pi} \left|\vec{{E}}
\right|^2~\mathrm{and}~~{w}_m = \frac{1}{8\pi} \left|\vec{{B}}
\right|^2~. \label{ST}
\end{eqnarray}
Eq. (\ref{PoyT}) is an exact identity following from Maxwell's
equations in the time domain. Quantities in Eq. (\ref{PoyT}) have
the dimension of [energy]/[time].

As is well-known \cite{JACK}, a version of the Poynting theorem in
the space-frequency domain can be derived starting with the
Poynting theorem in the space-time domain formulated for a
monochromatic field like in Eq. (\ref{wavemo}) and averaging over
a cycle of oscillation in time. Then, use of Eq. (\ref{rel0})
yields the following complex relation for space-frequency domain
quantities:

\begin{eqnarray}
2 \int_V \vec{\bar{j}}^{~*} \cdot \vec{\bar{E}} ~d V+ \int_A
\vec{\bar{S}}\cdot \hat{n}~ d A + i \omega \int_V
\left(\bar{w}_e-\bar{w}_m\right) ~d V =0~. \label{Poy1}
\end{eqnarray}
Here we define the complex, space-frequency domain version of the
Poynting vector $\vec{\bar{S}}$ and of the electric $\bar{w}_e$
and magnetic $\bar{w}_m$ energy densities in vacuum as

\begin{eqnarray}
\vec{\bar{S}} = \frac{c}{2\pi} \left(\vec{\bar{E}} \times
\vec{\bar{B}}^{~*}\right)~,~~~\bar{w}_e = \frac{1}{4\pi}
\left|\vec{\bar{E}} \right|^2~\mathrm{and}~~\bar{w}_m =
\frac{1}{4\pi} \left|\vec{\bar{B}} \right|^2~. \label{S}
\end{eqnarray}
Eq. (\ref{Poy1}) is an exact identity following from Maxwell's
equations in the frequency domain. Note that now quantities in Eq.
(\ref{Poy1}) have the dimension of [energy]/[frequency], i.e. of a
spectral energy density integrated through a given surface. In our
case, we can separately write real and imaginary parts of Eq.
(\ref{Poy1}) as

\begin{eqnarray}
2 \int_V \mathrm{Re}\left[\vec{\bar{j}}^{~*} \cdot
\vec{\bar{E}}\right] ~d V+ \int_A
\mathrm{Re}\left[\vec{\bar{S}}\cdot \hat{n}\right]~ d A =0~.
\label{Poy1R}
\end{eqnarray}
and

\begin{eqnarray}
2 \int_V \mathrm{Im}\left[\vec{\bar{j}}^{~*} \cdot
\vec{\bar{E}}\right] ~d V+  \omega \int_V
\left(\bar{w}_e-\bar{w}_m\right) ~d V + \int_A
\mathrm{Im}\left[\vec{\bar{S}}\cdot \hat{n}\right]~ d A=0~.
\label{Poy1I}
\end{eqnarray}
Eq. (\ref{Poy1R}) expresses energy conservation of time-averaged
quantities, while Eq. (\ref{Poy1I}) "relates to the reactive or
stored energy and its alternating flow" (cited from \cite{JACK}).
\begin{figure}
\begin{center}
\includegraphics*[width=140mm]{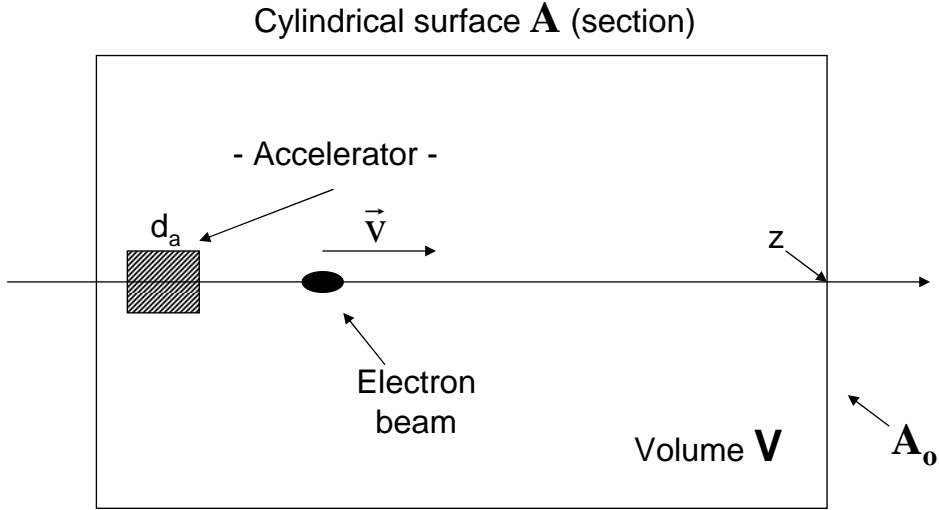}
\caption{\label{geom2} Geometry for application of Poynting
theorem.}
\end{center}
\end{figure}
In our case we consider the volume $V$ and the surface $A$ of
cylindrical shape as depicted in Fig. \ref{geom2}.  From Eq.
(\ref{Poy1R}) follows the equality between the total energy losses
of the whole bunch and the energy of coherent radiation in the far
zone, a fact that we will verify later on. Since $\vec{\bar{j}}$
is directed longitudinally we can write such equality (always with
reference to Fig. \ref{geom2}) as

\begin{eqnarray}
2 \int_V \mathrm{Re}\left[{\bar{j}}_z^{~*} {\bar{E}}_z\right] ~d V
= - \frac{c}{2\pi} \int_{A_o}
\left|\vec{\widetilde{E}}_\bot\right|^2~dA~, \label{repoy}
\end{eqnarray}
where $\vec{\widetilde{E}}_\bot$ indicates the radiation field.
The volume integral of the quantity $2 {\bar{j}}_z^{~*}
{\bar{E}}_z$ can be explicitly written with the help of Eq.
(\ref{solzo}) and Eq. (\ref{composj}) as

\begin{eqnarray}
2 \int_V  {\bar{j}}_z^{~*} {\bar{E}}_z ~d V =
\left|\bar{f}(\omega)\right|^2 X(\omega) \label{conx}
\end{eqnarray}
where

\begin{eqnarray}
X(\omega)&& = - \frac{2  \omega}{\gamma^2} \int d
\vec{r'}_{\bot}\int d \vec{r''}_{\bot} \rho^{~*}_o(\vec{r'}_\bot)
\rho_o(\vec{r''}_\bot) \int_{0}^{z} d z'  \cr && \times \Bigg\{i
\int_{0}^{z'} \frac{dz''}{(z'-z'')} \exp\left[\frac{i\omega
(z''-z')}{2 \gamma^2 c} + i\omega{\mid \vec{r}_{\bot o}
-\vec{r'}_\bot\mid^2\over{2c (z'-z'')}}\right] \cr &&
+\frac{\gamma^2 c}{\omega z'} \exp\left[ i\omega{\mid
\vec{r}_{\bot o} -\vec{r'}_\bot\mid^2\over{2c z'}}\right]
\exp\left[-\frac{i \omega z'}{2 \gamma^2 c}\right]\Bigg\}
~.\label{Xomega}
\end{eqnarray}
Note that that $\rho_o$ is a real quantity. The quantity
$X(\omega)$ has the property $X(\omega)=X^{~*}(-\omega)$, as it
can be directly verified. This means that $\mathrm{Re}[X](\omega)$
is an even function of $\omega$, while $\mathrm{Im}[X](\omega)$ is
odd. $X(\omega)$ is strictly related to $\Delta E_{\mathrm{tot}}$,
the total energy lost by the bunch as it travels within the volume
$V$. In fact, $\Delta E_{\mathrm{tot}}$ is given by

\begin{eqnarray}
\Delta \mathcal{E}_\mathrm{tot}= \frac{c}{4\pi} \int_{A_o} d A
\int_{-\infty}^{\infty} dt \left|\vec{E}(t)\right|^2
=\frac{c}{8\pi^2} \int_{A_o} d A \int_{-\infty}^{\infty} d\omega
\left|\vec{\widetilde{E}}_\bot\right|^2  ~.\label{elost0}
\end{eqnarray}
Comparison with Eq. (\ref{repoy}) yields

\begin{eqnarray}
\Delta \mathcal{E}_\mathrm{tot}=  \frac{1}{2\pi}
\int_{-\infty}^{\infty} d\omega \int_V
\mathrm{Re}\left[{\bar{j}}_z^{~*} {\bar{E}}_z\right] ~d V =
\frac{1}{4\pi} \int_{-\infty}^{\infty} d\omega
\left|\bar{f}(\omega)\right|^2 X(\omega) ~.\label{elost1}
\end{eqnarray}
Since $|\bar{f}|^2$ is even, $\mathrm{Re}[X](\omega)$ is even, and
$\mathrm{Im}[X](\omega)$ is odd, we conclude that $\Delta
\mathcal{E}_\mathrm{tot}$ is a positive real quantity, as it must
be.

As we will briefly review here, $X(\omega)$ is also strictly
related to the impedance of the system. When the bunch is ultra
relativistic, the longitudinal impedance of the system,
$Z_o(\omega)$ is typically given as the Fourier transform of the
wake function $G_o(\Delta s)$, that is

\begin{eqnarray}
Z_o(\omega) = \int_{-\infty}^{\infty} \frac{d(\Delta s)}{\beta c}~
G_o(\Delta s) \exp\left[i \omega \frac{\Delta s}{\beta c}\right]~,
\label{impe}
\end{eqnarray}
where the wake function is defined as
\begin{eqnarray}
G_o(\Delta s) = \frac{1}{(-e)} \int_{-\infty}^{\infty} d z'
~E_z(\Delta s, t)|_{ t = z'/(\beta c)}~. \label{wake2}
\end{eqnarray}
Here $E_z(\Delta s, t)$ indicates the longitudinal component of
the time-domain electric field generated by a source particle
acting on a test particle at longitudinal distance $\Delta s$ from
the source. This field is integrated along the test particle
trajectory, and divided by the electron charge $(-e)$, so that
$e^2 G_o(\Delta s)$ is the energy (gained, or lost) by the test
particle due to the action of the source. In agreement with
\cite{WAKK} we take the test particle behind the source for
positive values of $\Delta s$. According to the given definition
of wake function, one should integrate the longitudinal field over
the entire trajectory. However, there is no principle difficulty
in considering only part of the trajectory, let us say, up to
position $z$. Mathematically, this means that the upper
integration limit in Eq. (\ref{wake2}), i.e. $\infty$, should be
substituted with $z$. In this way, $G=G(\Delta s, z)$. Moreover,
our trajectory is supposed to start at $z_A=0$, that allows one to
substitute the lower integration limit in Eq. (\ref{wake2}), i.e.
$-\infty$, with $0$. We thus obtain

\begin{eqnarray}
G_o(\Delta s,z) = \frac{1}{(-e)} \int_{0}^{z} d z' ~E_z(\Delta s,
t)|_{ t = z'/(\beta c)}~. \label{wake1}
\end{eqnarray}
In both Eq. (\ref{impe}) and Eq. (\ref{wake1}) we considered a
case when test and source particles move along the longitudinal
$z$ axis. More in general, we should include given transverse
offsets of test and source with respect to the $z$ axis. Therefore
we have to modify Eq. (\ref{impe}) and Eq. (\ref{wake1}) to
include a dependence on such offset, i.e. on the test and source
transverse coordinates $\vec{r}_{\bot T}$ and $\vec{r}_{\bot S}$:
in this way $G_o = G_o(\Delta s,z,\vec{r}_{\bot S},\vec{r}_{\bot
T})$. In order to make our definitions independent of
$\vec{r}_{\bot T}$ and $\vec{r}_{\bot S}$, we integrate over the
transverse particle distribution in $d\vec{r}_{\bot T}$ and
$d\vec{r}_{\bot S}$. This can be interpreted as a substitution of
test and source particles with disks of charge longitudinally
separated by a distance $\Delta s$. We thus obtain

\begin{eqnarray}
G(\Delta s,z) = \frac{1}{e^2} \int d \vec{r'}_\bot \int d
\vec{r''}_\bot \rho_o(\vec{r'}_\bot) \rho_o(\vec{r''}_\bot)
G_o(\Delta s,z,\vec{r'}_{\bot},\vec{r''}_{\bot})~. \label{Gred}
\end{eqnarray}
In Eq. (\ref{Gred}) we used the fact that $\rho_o$ is independent
of the longitudinal position. With the redefinition in Eq.
(\ref{Gred}) we can further consider the impedance $Z(\omega,z)$:

\begin{eqnarray}
Z(\omega,z) = \int_{-\infty}^{\infty} \frac{d(\Delta s)}{\beta c}~
G(\Delta s,z) \exp\left[i \omega \frac{\Delta s}{\beta c}\right]~.
\label{impered}
\end{eqnarray}
As is well-known, since $G(\Delta s,z)$ is a real function one has
$Z(\omega,z) = Z^{~*}(-\omega,z)$, i.e. the impedance has even
real part and odd imaginary part. Thus, if we split the wake
function in the sum $G = G_S + G_A$ of a symmetric and
antisymmetric part, the Fourier transform of $G_S$ gives back
$\mathrm{Re}[Z]$, while the Fourier Transform of $G_A$ yields
$\mathrm{Im}[Z]$. The symmetric part of the wake, $G_S$, is also
called the active part and is related with the energy lost by the
bunch through radiation. The antisymmetric part $G_A$ instead, is
called the reactive part and is related with energy redistribution
within the bunch, but not with the energy radiated. This can be
seen by writing the total energy lost by the bunch in terms of
$G(\tau)$, where $\tau = s/(\beta c)$:

\begin{eqnarray}
\Delta \mathcal{E}_\mathrm{tot}(z) = \int_{-\infty}^{\infty} d t
\int_{-\infty}^{\infty} d \tau  ~ G(\tau,z) f(t) f(t- \tau) =
\int_{-\infty}^{\infty} d \tau  ~ G(\tau,z)
\mathcal{A}[f](\tau)~,\label{losene}
\end{eqnarray}
where $\mathcal{A}[f](\tau)$ indicates the autocorrelation
function of $f$. Since $\mathcal{A}[f](\tau)$ is even in $\tau$,
only the symmetric part $G_S$ enters effectively in the expression
for $\Delta \mathcal{E}_\mathrm{tot}$. Finally, using the relation

\begin{eqnarray}
\int_{-\infty}^{\infty} dt g_1(t) g_2^{~*}(t) = \frac{1}{2\pi}
\int_{-\infty}^{\infty} d \omega \bar{g}_1(\omega)
\bar{g}_2^{~*}(\omega) ~\label{parsesim}
\end{eqnarray}
to simplify Eq. (\ref{losene}) we can write

\begin{eqnarray}
\Delta \mathcal{E}_\mathrm{tot}(z) = \frac{1}{2\pi}
\int_{-\infty}^{\infty} d \omega~ Z(\omega,z)
\left|\bar{f}(\omega)\right|^2 ~,\label{losene2}
\end{eqnarray}
where we used the autocorrelation theorem to find
$\overline{\mathcal{A}[f]} = |\bar{f}|^2 =
(~\overline{\mathcal{A}[f]}~)^{~*}$ and Eq. (\ref{impered}). In
the case $|f(\omega)|^2 = e^2$ (single electron), by definition of
impedance, Eq. (\ref{impe}), we have

\begin{eqnarray}
Z(\omega,z) = \frac{1}{e^2} \int_V  {\bar{j}}_z^{~*} {\bar{E}}_z
~d V \label{defZZ}
\end{eqnarray}
and from Eq. (\ref{conx}) we have

\begin{eqnarray}
Z(\omega,z) = \frac{1}{2} X(\omega,z)~, \label{ZX}
\end{eqnarray}
that yields a practical algorithm to compute both real and
imaginary parts of the impedance (and, subsequently, of the wake),
through the evaluation of Eq. (\ref{Xomega}).

Finally, a word of caution. The impedance formalism is often used
in order to describe coupling impedances. A typical situation is
the following.  A given electron in a bunch induces charge and
current densities in the vacuum chamber walls. In turn, these
charge and current densities are responsible for spurious
electromagnetic fields acting back on electrons following the
exciting one, thus generating a coupling wake function (and thus
its related coupling impedance). This coupling wake function $G_c$
must vanish in front of an exciting particle in the limit for
$\gamma \longrightarrow \infty$, i.e. for electron velocities
$v\longrightarrow c$. As a result one can take $G_c(\Delta s) = 0$
for $\Delta s < 0$ (see \cite{WAKK}). This property is necessary
(although not sufficient) to demonstrate that the coupling
impedance $Z_c$ (i.e. the Fourier transform of $G_c$), obeys the
well-known Kramers-Kronig relation \cite{KRAM,KRON}, thus
providing a link between real and imaginary parts of the
impedance. However, our impedance does not obey Kramers-Kronig
relations. In fact, we are dealing with a mechanism of generation
of the wake function that is fundamentally different from that of
coupling wakes. In particular $G(\Delta s)$ can be different from
zero both for $\Delta s<0$ and $\Delta s>0$, a result of the fact
that we are dealing both with tail-head as well as with head-tail
interactions in the time domain\footnote{Note that causality with
respect to time remains obviously valid but it cannot be
exploited, as in the case of coupling wakes, giving a relation
between real and imaginary part of the impedance.}.

\section{\label{sec:asymp} Analytical asymptotes of the longitudinal
impedance}

Let us write an explicit expression for the impedance with the
help of Eq. (\ref{Xomega}):

\begin{eqnarray}
Z(\omega,z) = Z_1(\omega,z) + Z_2(\omega,z)~.\label{Zexpl}
\end{eqnarray}
Here we have defined, for computational convenience, the
quantities

\begin{eqnarray}
Z_1(\omega,z)&& = - \frac{i \omega}{\gamma^2} \int d
\vec{r'}_{\bot}\int d \vec{r''}_{\bot} \rho^{~*}_o(\vec{r'}_\bot)
\rho_o(\vec{r''}_\bot) \cr && \times  \int_{0}^{z} d z'
\int_{0}^{z'} \frac{dz''}{(z'-z'')} \exp\left[\frac{i\omega
(z''-z')}{2 \gamma^2 c} + i\omega{\mid \vec{r'}_{\bot}
-\vec{r''}_\bot\mid^2\over{2c (z'-z'')}}\right] ~,\label{Zexpl1}
\end{eqnarray}
and

\begin{eqnarray}
Z_2(\omega,z)&& = -  \int d \vec{r'}_{\bot}\int d \vec{r''}_{\bot}
\rho^{~*}_o(\vec{r'}_\bot) \rho_o(\vec{r''}_\bot) \cr && \times
\int_{0}^{z} d z' \frac{c}{z'} \exp\left[ i\omega{\mid
\vec{r'}_{\bot} -\vec{r''}_\bot\mid^2\over{2c
z'}}\right]\exp\left[-\frac{i \omega z'}{2 \gamma^2
c}\right]~,\label{Zexpl2}
\end{eqnarray}
Aside for particular cases, the task of calculating real and
imaginary part of the impedance must be performed numerically. It
is possible, however, to obtain important analytical information
about the impedance and wake function in the asymptotic case $z
\gg 2 \gamma^2 \lambdabar$ from Eq. (\ref{Zexpl}).

\subsection{\label{sub:re} Real part}

Let us consider $\mathrm{Re}[Z]$ for $z \gg 2 \gamma^2
\lambdabar$. We first deal with $Z_2$. With the help of Eq.
(\ref{rele}), it can be shown that in the limit for $z \gg 2
\gamma^2 \lambdabar$, $Z_2$ tends to a real quantity independent
of $z$:

\begin{eqnarray}
Z_2(\omega)&& = - {2 c} \int d \vec{r'}_{\bot}\int d
\vec{r''}_{\bot} \rho^{~*}_o(\vec{r'}_\bot) \rho_o(\vec{r''}_\bot)
 K_o\left(\frac{\omega \mid
\vec{r'}_{\bot} -\vec{r''}_\bot\mid}{\gamma
c}\right)~.\label{Zexpl2re}
\end{eqnarray}
The integral in $d \vec{r''}_\bot$ can be interpreted as a
convolution product between two functions of $\vec{r}_\bot$, i.e.
$\rho_o$ and $K_o$. This is equal to the anti-Fourier transform in
two dimensions of the product of the spatial Fourier transforms of
$\rho_o$ and $K_o$, that we will call respectively
$\tilde{\rho}_o(\vec{k})$ and $\tilde{K}_o(\vec{k})$, $\vec{k}$
being the conjugate variable to $\vec{r}_\bot$. By exchanging the
integration in $d \vec{k}$ and that in $d\vec{r'}_\bot$ one
obtains the following alternative representation of $Z_2$:

\begin{eqnarray}
Z_2(\omega)&& = - \frac{c}{2 \pi^2} \int d \vec{k}
\left|\tilde{\rho}_o(\vec{k})\right|^2
\tilde{K}_o\left(\vec{k}\right)~.\label{Z2ref}
\end{eqnarray}
As concerns $Z_1$ we make use of the two identities

\begin{eqnarray}
&&\frac{2\pi i c}{\omega} \frac{z'z''}{z'-z''} \exp\left[-\frac{i
\omega | \vec{r'}_{\bot} -\vec{r''}_\bot|^2 }{2 c (z'-z'')}\right]
= \cr &&    \int d \vec{r}_{\bot} \exp\left[-\frac{i \omega |
\vec{r}_{\bot} -\vec{r'}_\bot|^2 }{2 c z'}+\frac{i \omega |
\vec{r}_{\bot} -\vec{r''}_\bot|^2 }{2 c z''}\right] ~\label{id1}
\end{eqnarray}
and

\begin{eqnarray}
\int_{0}^{z} g(z')~ dz' \int_{0}^{z'} g^*(z'')~ dz'' +C.C. =
\left|\int_{0}^{z} g(z)~ dz ~\right|^2 ~,\label{id2}
\end{eqnarray}
and of Eq. (\ref{rele}). With the help of these relations we
rewrite the real part of $Z_1$ in the limit for $z \gg 2
\gamma^2\lambdabar$ as an expression independent of $z$:

\begin{eqnarray}
\mathrm{Re}[Z_1](\omega)&& = \frac{ \omega^2}{ \pi \gamma^2 c }
\int d \vec{r}_{\bot} \left| \int d \vec{r'}_{\bot}
\rho_o(\vec{r'}_\bot) K_o\left(\frac{\omega \mid \vec{r}_{\bot }
-\vec{r'}_\bot\mid}{\gamma c}\right)\right|^2~.\label{Zexpl1v1}
\end{eqnarray}
Applying Eq. (\ref{parsesim}) with $g_1=g_2$ to Eq.
(\ref{Zexpl1v1}) and remembering that the spatial Fourier
transform of the convolution product of two function is equal to
the product of the spatial Fourier transform of the same function,
we can rewrite Eq. (\ref{Zexpl1v1}) as

\begin{eqnarray}
\mathrm{Re}[Z_1](\omega)&& = \frac{\omega^2}{4 \pi^3 \gamma^2}
\int d \vec{k} \left| \tilde{\rho}_o(\vec{k})
\tilde{K}_o\left(\vec{k}\right)\right|^2~.\label{Z1ref}
\end{eqnarray}
Thus, putting together Eq. (\ref{Z2ref}) and Eq. (\ref{Z1ref}) we
obtain

\begin{eqnarray}
\mathrm{Re}[Z](\omega)&& = \frac{\omega^2}{4 \pi^3 \gamma^2} \int
d \vec{k} \left| \tilde{\rho}_o(\vec{k})
\tilde{K}_o\left(\vec{k}\right)\right|^2- \frac{  c}{2  \pi^2}
\int d \vec{k} \left|\tilde{\rho}_o(\vec{k})\right|^2
\tilde{K}_o\left(\vec{k}\right)~.\label{Zretot}
\end{eqnarray}
Substitution in Eq. (\ref{Zretot}) of the explicit expression for
$\tilde{K}_o(\vec{k})$

\begin{eqnarray}
\tilde{K}_o\left(\vec{k}\right) &=&\int d\vec{r}_\bot \exp\left[i
\vec{k}\cdot \vec{r}_\bot\right] K_o\left(\frac{\omega
\left|\vec{r}_\bot\right|}{\gamma c}\right) = 2\pi \int_0^\infty
dr_\bot r_\bot K_o\left(\frac{\omega r_\bot}{\gamma c}\right)
J_o(k r_\bot) \cr &&= \frac{2\pi c^2 \gamma^2}{\omega^2}
\frac{1}{\left[1+\left(c\gamma k/\omega\right)^2\right]}~,
\label{kok}
\end{eqnarray}
together with the new notation $\vec{\theta} = c \vec{k}/\omega$
yields for the sum of two integrals in Eq. (\ref{Zretot}):

\begin{eqnarray}
\mathrm{Re}[Z](\omega) = - \frac{c}{\pi} \int d \vec{\theta}
\left| \tilde{\rho}_o(\vec{\theta},\omega)\right|^2
\frac{\gamma^4\theta^2}{\left(1+\gamma^2\theta^2\right)^2}
~.\label{Zre}
\end{eqnarray}
Up to now we did not make use of any particular model for the
electron beam. Choosing Eq. (\ref{rhor}) as a model for $\rho_o$
we can substitute into Eq. (\ref{Zre}) the following expression
for $\tilde{\rho}_o$:

\begin{eqnarray}
\tilde{\rho}_o\left(\vec{\theta},\omega\right) = \frac{1}{c}
\exp\left[-\frac{\theta^2 \omega^2 \sigma_\bot^2}{2 c^2}\right]~.
\label{rhotrasf}
\end{eqnarray}

\begin{figure}
\begin{center}
\includegraphics*[width=140mm]{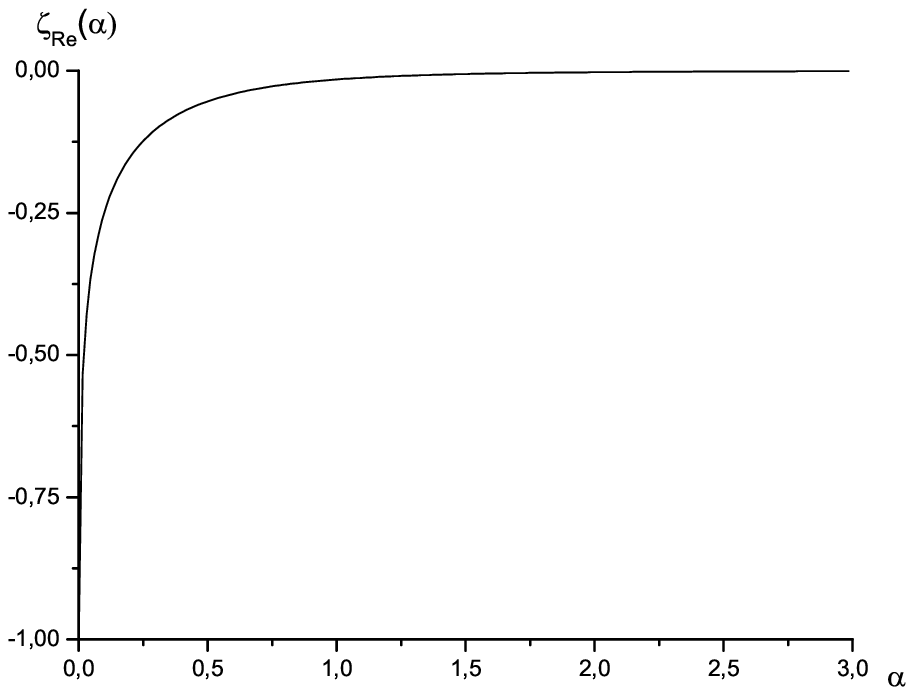}
\caption{\label{rez} Plot of the universal function
$\zeta_{\mathrm{Re}}$, as a function of $\alpha$. }
\end{center}
\end{figure}
This gives

\begin{eqnarray}
\mathrm{Re}[Z] = Z_o \zeta_{\mathrm{Re}}(\alpha) ~,\label{Zrexpl}
\end{eqnarray}
where

\begin{eqnarray}
\zeta_{\mathrm{Re}}(\alpha) = \frac{1}{4\pi}\left\{1-(1+\alpha^2)
\exp[\alpha^2] \Gamma(0,\alpha^2) \right\} ~,\label{zetaim}
\end{eqnarray}
the dimensionless parameter $\alpha$ is defined as

\begin{eqnarray}
\alpha = \frac{\omega \sigma_\bot}{\gamma c} ~,\label{alphad}
\end{eqnarray}
and

\begin{eqnarray}
\Gamma(0,x) = \int_x^\infty \frac{dt}{t} \exp[-t]~ \label{Gdef}
\end{eqnarray}
and $Z_o = 4\pi/c$ indicates the free-space impedance. A plot of
the universal function $\zeta_{\mathrm{Re}}(\alpha)$, that is the
real part of the impedance in units of $Z_o$, is given in Fig.
\ref{rez}. Note that $\mathrm{Re}[Z]$ exhibits a singular behavior
for $\sigma_\bot \longrightarrow 0$ (i.e. for $\alpha
\longrightarrow 0$ in Fig. \ref{rez}). This is linked with our
particular model, where we chose $d_a \longrightarrow 0$.

\subsection{\label{sub:im} Imaginary part}

Since, in the limit for $z \gg \gamma^2 \lambdabar$, $Z_2$ is a
real quantity, from Eq. (\ref{Zexpl1}) follows that  the imaginary
part of $Z$ is

\begin{eqnarray}
\mathrm{Im}[Z](\omega,z)&& = - \frac{\omega}{2 \gamma^2}
~\mathrm{Im}\Bigg\{\int d \vec{r'}_{\bot}\int d \vec{r''}_{\bot}
\rho^{~*}_o(\vec{r'}_\bot) \rho_o(\vec{r''}_\bot) \cr && \times
\int_{0}^{z} d z' \int_{0}^{z'} dz''\frac{2 i}{(z'-z'')}
\exp\left[\frac{i\omega (z''-z')}{2 \gamma^2 c} + i\omega{\mid
\vec{r'}_{\bot} -\vec{r''}_\bot\mid^2\over{2c (z'-z'')}}\right]
\Bigg\} .\cr &&\label{ImZlim}
\end{eqnarray}
In the asymptotic limit for $z \gg 2 \gamma^2 \lambdabar$ one
finds the same result that one would have found calculating the
impedance with the steady state expression for the field, Eq.
(\ref{solstdy}), and substituting $z_o$ with zero that is

\begin{eqnarray}
\mathrm{Im}[Z](\omega,z)&& = - \frac{\omega}{2 \gamma^2}
~\mathrm{Im}\Bigg\{\int d \vec{r'}_{\bot}\int d \vec{r''}_{\bot}
\rho^{~*}_o(\vec{r'}_\bot) \rho_o(\vec{r''}_\bot) \cr && \times
\int_{0}^{z} d z' \int_{0}^{\infty} dz''\frac{2 i}{(z'-z'')}
\exp\left[\frac{i\omega (z''-z')}{2 \gamma^2 c} + i\omega{\mid
\vec{r'}_{\bot} -\vec{r''}_\bot\mid^2\over{2c (z'-z'')}}\right]
\Bigg\} .\cr &&\label{Zim1}
\end{eqnarray}
With the help of Eq. (\ref{rele}) we find the analogous of Eq.
(\ref{Zexpl1v1}) for the imaginary part of the impedance in the
asymptotic limit for $z \gg 2 \gamma^2 \lambdabar$:

\begin{eqnarray}
\mathrm{Im}[Z](\omega,z)&& = - \frac{2 \omega z}{\gamma^2} ~ \int
d \vec{r'}_{\bot}\int d \vec{r''}_{\bot}
\rho^{~*}_o(\vec{r'}_\bot) \rho_o(\vec{r''}_\bot)
K_o\left(\frac{\omega \mid \vec{r}_{\bot }
-\vec{r'}_\bot\mid}{\gamma c}\right)~.\label{Zimtot1}
\end{eqnarray}
Similarly as for Eq. (\ref{Zexpl2re}), we observe that the
integral in $\vec{r''}_\bot$ can be interpreted as a convolution.
In analogy with Eq. (\ref{Z2ref}) we obtain

\begin{eqnarray}
\mathrm{Im}[Z](\omega,z)&& = - \frac{  \omega z}{2\pi^2 \gamma^2}
\int d \vec{k} \left|\tilde{\rho}_o(\vec{k})\right|^2
\tilde{K}_o\left(\vec{k}\right)~.\label{Zimref}
\end{eqnarray}
Finally, Eq. (\ref{kok}) together with the notation $\vec{\theta}
= c \vec{k}/\omega$ yields

\begin{eqnarray}
\mathrm{Im}[Z](\omega) = - \frac{1}{\pi} \omega z \int d
\vec{\theta} \left| \tilde{\rho}_o(\vec{\theta},\omega)\right|^2
\frac{1}{\left(1+\gamma^2\theta^2\right)} ~.\label{Zim}
\end{eqnarray}
Similarly as for the real part, use of Eq. (\ref{rhotrasf}) allows
to give an explicit expression for $\mathrm{Im}[Z]$:

\begin{figure}
\begin{center}
\includegraphics*[width=140mm]{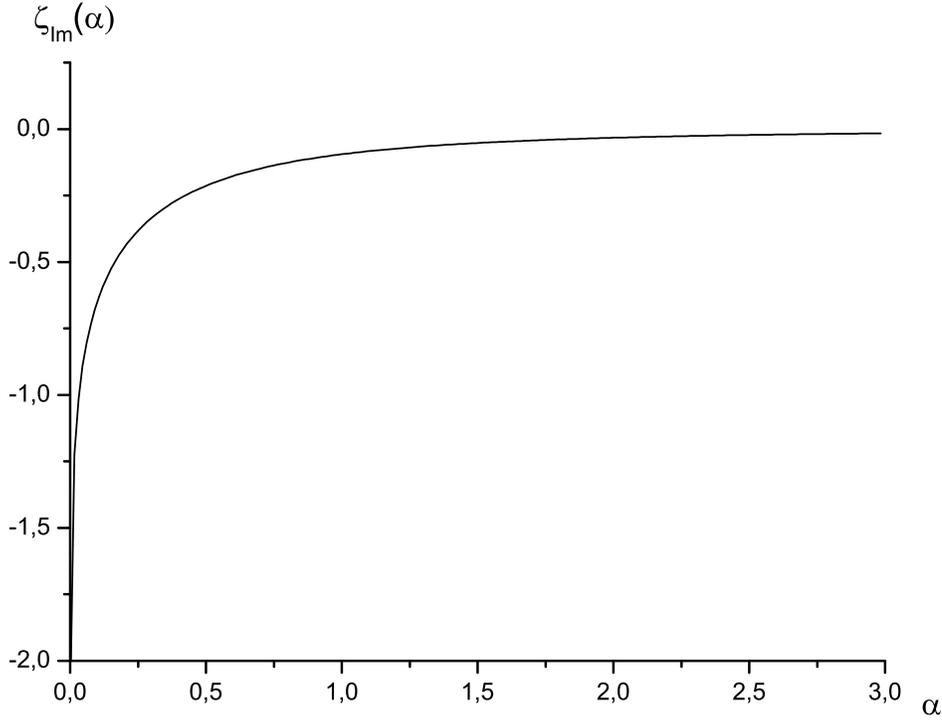}
\caption{\label{imz} Plot of the universal function
$\zeta_{\mathrm{Im}}$, as a function of $\alpha$. }
\end{center}
\end{figure}

\begin{eqnarray}
\mathrm{Im}[Z] = (Z_o \hat{z}) ~\zeta_{\mathrm{Im}}(\alpha)
~,\label{Zimexpl}
\end{eqnarray}
where

\begin{eqnarray}
\zeta_{\mathrm{Im}}(\alpha) = -\frac{1}{2\pi} \exp[\alpha^2]
~\Gamma(0,\alpha^2)  ~,\label{zetaim2}
\end{eqnarray}
the dimensionless parameter $\alpha$ is defined as in Eq.
(\ref{alphad}), $\Gamma$ was given in Eq. (\ref{Gdef}), $Z_o =
4\pi/c$ indicates, as before, the free-space impedance, and
$\hat{z} = z/(2\gamma^2 \lambdabar)$. A  plot of the universal
function $\zeta_{\mathrm{Im}}(\alpha)$, that is the imaginary part
of the impedance in units of $Z_o \hat{z}$ is given in Fig.
\ref{imz}.

Note that one may use different models for the electron beam
transverse profile to be substituted in Eq. (\ref{Zim}). For
example, one may consider a stepped profile defined by

\begin{eqnarray}
\rho_o(\vec{r}_\bot) = \frac{1}{\pi r_b^2 c}~
\mathcal{C}_{r_b}(\vec{r}_\bot)~, \label{rhostep}
\end{eqnarray}
where $r_b$ is now the beam radius and
$\mathcal{C}_{r_b}(\vec{r}_\bot)$ is a circle function, equal to
unity for $|\vec{r}_\bot|<r_b$ and zero otherwise. The analogous
of Eq. (\ref{rhotrasf}) is

\begin{eqnarray}
\tilde{\rho}_o(\vec{\theta},\omega) = \frac{2}{\omega r_b \theta}
J_1\left(\frac{\omega r_b \theta}{c}\right) ~. \label{rhosttras}
\end{eqnarray}
By substituting Eq. (\ref{rhosttras}) in Eq. (\ref{Zim}) and
performing integration we recover the already known expression for
the longitudinal space-charge impedance in case of a stepped
profile bunch

\begin{eqnarray}
Z_{LSC}(\omega) = -\frac{4 i c z}{\omega
r_b^2}\left[1-\frac{\omega r_b }{c \gamma} K_1\left(\frac{\omega
r_b}{\gamma c}\right)\right], \label{ZLSC}
\end{eqnarray}
in agreement with \cite{SHAF} where an expression per unit length
is given, and the overall difference of a minus sign fixes the
sign of the energy loss (using the convention in Eq. (\ref{ZLSC})
it is negative).

\section{\label{sec:conse} Energy conservation}

It is interesting to verify the energy conservation law for the
active part of the field, that is Eq. (\ref{repoy}).

With the help of Eq. (\ref{conx}), Eq. (\ref{ZX})  and Eq.
(\ref{Zre}) we can immediately write the left hand side of Eq.
(\ref{repoy}) in its final form as

\begin{eqnarray}
2 \int_V \mathrm{Re}\left[{\bar{j}}_z^{~*} {\bar{E}}_z\right] ~d V
= - \frac{2 c}{\pi} \left| \bar{f}(\omega) \right|^2  \int d
\vec{\theta} \left| \tilde{\rho}_o(\vec{\theta},\omega)\right|^2
\frac{\gamma^4\theta^2}{\left(1+\gamma^2\theta^2\right)^2} ~.
\label{repoyleft}
\end{eqnarray}
We now need to calculate the right hand side of Eq. (\ref{repoy}).
This is the spectral energy density of radiation integrated over a
transverse plane in the far zone, i.e. at a large longitudinal
distance $z$ with respect to the radiation formation length $2
\gamma^2 \lambdabar$ at all wavelengths of interest, i.e. for
$\lambdabar \gtrsim \sigma_z$. This gives $z \gg 2\gamma^2
\sigma_z$, $2 \gamma^2 \sigma_z$ being the already-defined
overtaking length. Radiation is generated because of the
acceleration process, and is linked with the longitudinal wakes
through Eq. (\ref{repoy}). Radiation and longitudinal wakes should
actually be seen as two faces of the same coin. We have already
seen that major simplifications arise in the calculation of the
impedance (and, therefore, of the wake) as one assumes a very
short acceleration distance $d_a$ compared with the overtaking
length, $d_a \ll 2 \gamma^2 \sigma_z$. In particular, in this
case, the impedance does not depend on the details of the
acceleration process, that can be considered as taking place at a
single point. This fact has its reflection on the characteristics
of radiation, that are also bound to be independent of how
acceleration took place, always provided that $d_a \ll 2
\gamma^2\lambdabar$ at all wavelengths of interest. The
acceleration process works as a switch in the space-frequency
domain, in the sense that it switches on the harmonic contents of
the electromagnetic sources at a given wavelength of interest. As
long as it takes place within a distance much smaller than
$2\gamma^2 \lambdabar$, such switching process may have very
different physical realizations. It may be due to an ultra-high
field gradient, but also to a bending magnet deflecting an
electron beam from an off-axis trajectory to a trajectory along
the $z$ axis, or to the crossing of an interface between two media
with different dielectric properties. All these examples produce
well-documented types of collimated radiation in the long
wavelength range ($\lambdabar \gtrsim \sigma_z$). Radiation due to
longitudinal acceleration is studied in \cite{JACK}, while if the
switcher is a bend one obtains edge radiation (see \cite{BOS4}
among many others, and references therein). Finally, passage
through an interface between different media produces transition
radiation. In particular, coherent transition radiation in the far
zone generated at the interface between plasma and vacuum in a
laser-plasma accelerator (with $d_a \ll 2\gamma^2 \lambdabar$) has
first been studied in \cite{SCHR}. Characteristics of these kinds
of radiation coincide as long as the switching process obeys $d_a
\ll 2\gamma^2 \lambdabar$ \footnote{From this viewpoint, we do not
agree with \cite{SCHR} about the presence of diffraction radiation
when the electron beam is created inside the plasma. There is a
principle difference between a case when an electron travels
through a foil with finite transverse size and another when an
electron is created at the interface between plasma and vacuum.
When $d_a \ll 2\gamma^2 \lambdabar$ radiation characteristics
should not depend on the transverse characteristics of the
plasma.}.

In reference \cite{OURF} we showed how radiation produced by a
single ultrarelativistic electron can be suggestively described in
terms of a laser-like beam. Once the waist of this laser beam is
specified, the field distribution at any position down the
$z$-axis can be found applying Fourier Optics techniques (in
free-space this amounts to the application of Fourier propagation
equation). In particular, a single electron created at $z_A=0$ by
the system depicted in Fig. \ref{geom2} produces a laser-like beam
is with a waist located at $z_A = 0$. As we have just discussed,
since $d_a \ll 2 \gamma^2 \lambdabar$, the field distribution at
the waist does not depend on the realization of the accelerator
setup, and coincides with that produced by a magnet edge, or by
transition radiation. It is given by the field distribution
associated with a single electron created at $z_A=0$ and
travelling along the $z$ axis \cite{OURF}:

\begin{eqnarray}
\vec{\widetilde{E}}_{\bot s}(\vec{r}_\bot) = \frac{2 i (-e)
\omega}{c^2 \gamma} \frac{\vec{r}_\bot}{r_\bot} K_1
\left(\frac{\omega r_\bot}{c\gamma}\right)~. \label{vsour}
\end{eqnarray}
In our case though, we are considering an electron beam and not a
single electron. This means that we are not dealing with a single
laser-like beam but, rather, with a coherent collection of
laser-like beams. At the initial point $z_A=0$ each electron is
completely characterized by an offset $\vec{r'}_\bot$ and a
deflection angle $\vec{\phi'}_\bot$ with respect to the $z$ axis.
These two vectors identify a point in the transverse electron
phase space. Each electron, characterized by the pair
$(\vec{r'}_\bot,\vec{\phi'}_\bot)$, corresponds to a different
laser-like beam, whose waist is simply tilted of an angle
$\vec{\phi'}_\bot$ and shifted of an offset $\vec{r'}_\bot$.
Accounting for offsets and deflections, Eq. (\ref{vsour}) can be
written as

\begin{eqnarray}
\vec{\widetilde{E}}_{\bot s}(\vec{r}_\bot) = \frac{2 i (-e)
\omega}{c^2 \gamma} \exp\left[i \frac{\omega}{c}
\vec{\phi'}\cdot\left(\vec{r}_\bot-\vec{r'}_\bot\right)\right]
\frac{\vec{r}_\bot-\vec{r'}_\bot}{\left|\vec{r}_\bot-\vec{r'}_\bot\right|}
K_1 \left(\frac{\omega
\left|\vec{r}_\bot-\vec{r'}_\bot\right|}{c\gamma}\right)~.
\label{vsour21}
\end{eqnarray}
We now have to average Eq. (\ref{vsour21}) over the harmonic
component of the charge density at frequency $\omega$, i.e.
$\bar{\rho}$ in Eq. (\ref{composrho}) calculated at $z'=0$, at the
waist position. In principle,  $\bar{\rho}$ should depend on both
$\vec{\phi'}_\bot$ and $\vec{r'}_\bot$. However, in our model, it
only depends on $\vec{r'}_\bot$, since we posed $\rho_o =
\rho_o(\vec{r'}_\bot)$. In our situation of interest we can set
$\vec{\phi'}_\bot=0$ in Eq. (\ref{vsour21}) and obtain the
following average over $\bar{\rho}$:

\begin{eqnarray}
\vec{\widetilde{E}}_{\bot}(\vec{r}_\bot) = - \frac{2 i \omega}{c^2
\gamma} \bar{f}(\omega) \int d\vec{r'}_\bot
{\rho_o}(\vec{r'}_\bot)
\frac{\vec{r}_\bot-\vec{r'}_\bot}{\left|\vec{r}_\bot-\vec{r'}_\bot\right|}
K_1 \left(\frac{\omega
\left|\vec{r}_\bot-\vec{r'}_\bot\right|}{c\gamma}\right)~.
\label{vsour2}
\end{eqnarray}
The right hand side of Eq. (\ref{repoy}) in the space-frequency
domain, is given by calculating the flux of the complex Poynting
vector through the surface $A_o$ (shown in Fig. \ref{geom2}). Such
flux is equal to the flux of the complex Poynting vector
associated with the virtual source through the virtual source,
because the Fresnel propagator conserves the flux of the Poynting
vector. Therefore the right hand side of Eq. (\ref{repoy}) amounts
to

\begin{eqnarray}
&& - \frac{c}{2\pi} \int_{A_o}
\left|\vec{\widetilde{E}}_\bot\right|^2~ d A = \cr && -
\frac{c}{2\pi} \int d\vec{r}_\bot \left|\frac{2 \omega}{c^2
\gamma} \bar{f}(\omega) \int d\vec{r'}_\bot
{\rho_o}(\vec{r'}_\bot)
\frac{\vec{r}_\bot-\vec{r'}_\bot}{\left|\vec{r}_\bot-\vec{r'}_\bot\right|}
K_1 \left(\frac{\omega
\left|\vec{r}_\bot-\vec{r'}_\bot\right|}{c\gamma}\right)\right|^2~.
\label{repoyright}
\end{eqnarray}
Note that the integral in $d \vec{r'}_\bot$ is a convolution
product. As before, applying Eq. (\ref{parsesim}) with $g_1=g_2$
to Eq. (\ref{repoyright}) and remembering that the spatial Fourier
transform of the convolution product of two function is equal to
the product of the spatial Fourier transform of the same
functions, we can rewrite Eq. (\ref{repoyright}) as

\begin{eqnarray}
- \frac{c}{2\pi} \int_{A_o}
\left|\vec{\widetilde{E}}_\bot\right|^2~ d A = - \frac{2 \omega^2
}{\pi c \gamma^2} \frac{1}{4\pi^2} \left|\bar{f}(\omega)\right|^2
\int d\vec{k} \left|\tilde{\rho}_o(\vec{k})
\tilde{\mathcal{F}}\left(\vec{k}\right)\right|^2~,
\label{repoyright2}
\end{eqnarray}
where

\begin{eqnarray}
\tilde{\mathcal{F}}\left(\vec{k}\right) = \int d \vec{r'}_\bot
\exp\left[i \vec{k} \cdot \vec{r}_\bot\right]
\frac{\vec{r}_\bot}{r_\bot} K_1\left(\frac{\omega
r_\bot}{c\gamma}\right) = \frac{2\pi \gamma^2 c^2}{\omega^2}
\frac{\gamma \theta}{1+\gamma^2\theta^2}~, \label{fourdef}
\end{eqnarray}
having used the notation $\theta = c/\omega \vec{k}$. Substituting
Eq. (\ref{fourdef}) into Eq. (\ref{repoyright2}) we obtain the
final result

\begin{eqnarray}
- \frac{c}{2\pi} \int_{A_o}
\left|\vec{\widetilde{E}}_\bot\right|^2~ d A = - \frac{2 c }{\pi}
\left|\bar{f}(\omega)\right|^2 \int d\vec{\theta}
\left|\tilde{\rho}_o(\vec{\theta},\omega)\right|^2 \frac{\gamma^4
\theta^2}{\left(1+\gamma^2\theta^2\right)^2}~, \label{repoyright3}
\end{eqnarray}
that coincides with the left hand side in Eq. (\ref{repoyleft})
thus verifying Eq. (\ref{repoy}) as it must be.

\section{\label{sec:enchange} Analytical asymptote of the wake function}

The anti-Fourier transform of the impedance gives back the wake
function, that in its turn allows one to calculate the energy
change per particle averaged over the transverse beam coordinates.

Analytical results can be found starting with the asymptotic limit
of the impedance for $z \gg \gamma^2 \lambdabar$.

\subsection{\label{sub:symm} Symmetric part of the wake}

\begin{figure}
\begin{center}
\includegraphics*[width=140mm]{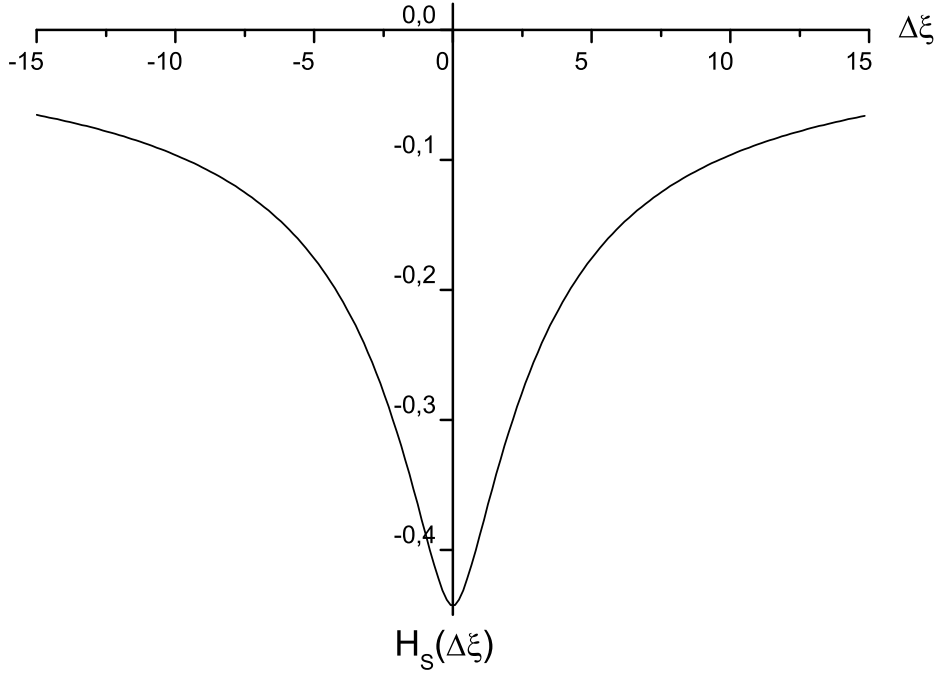}
\caption{\label{gs} Plot of the universal function $H_S$ as a
function of $\Delta \xi$. }
\end{center}
\end{figure}

The symmetric part of the wake function $G_S$ can be found
calculating

\begin{eqnarray}
G_S(\Delta s) = \frac{1}{2\pi} \int_{-\infty}^{\infty} d\omega~
\mathrm{Re}[Z](\omega) \exp\left[-i \omega \left(\frac{\Delta
s}{\beta c}\right)\right]~. \label{sym1}
\end{eqnarray}
With the help of Eq. (\ref{Zre}) and Eq. (\ref{rhotrasf}) one can
rewrite Eq. (\ref{sym1}) as

\begin{eqnarray}
G_S(\Delta s) &=& -  \frac{1}{\pi c} \int_0^{\infty} d {\theta}
\frac{\gamma^4\theta^3} {\left(1+\gamma^2\theta^2\right)^2}
\int_{-\infty}^{\infty} d\omega~ \exp\left[-\frac{\theta^2
\omega^2 \sigma_\bot^2}{c^2}\right] \exp\left[-i \omega
\left(\frac{\Delta s}{c}\right)\right]~.\cr && \label{sym2}
\end{eqnarray}
We first calculate the integral in $d \omega$, thus obtaining

\begin{eqnarray}
G_S(\Delta s) = -  \frac{1}{ \sqrt{\pi} \sigma_\bot}
\int_0^{\infty} d {\theta} \frac{\gamma^4\theta^2}
{\left(1+\gamma^2\theta^2\right)^2} \exp\left[-\frac{(\Delta
s)^2}{4 \theta^2 \sigma_\bot^2} \right]~. \label{sym3}
\end{eqnarray}
Performing the integral in $d\theta$ and using the notation
$\Delta \xi = \gamma (\Delta s)/\sigma_\bot$ one obtains

\begin{eqnarray}
G_S(\Delta \xi) = \frac{\gamma}{\sigma_\bot} H_S(\Delta \xi)
\label{sym4}
\end{eqnarray}
where

\begin{eqnarray}
H_S(\Delta \xi) = - \frac{1}{4\sqrt{\pi}}\left\{\sqrt{\pi}~
|\Delta \xi| + \pi \left[1-\frac{(\Delta\xi)^2}{2}\right]
\exp\left[\frac{(\Delta\xi)^2}{4}\right]
\mathrm{erfc}\left[\frac{|\Delta\xi|}{2}\right]\right\}~,
\label{HS}
\end{eqnarray}
and

\begin{eqnarray}
\mathrm{erfc}(x) = 1 - \frac{2}{\sqrt{\pi}} \int_0^x \exp[-t^2]
dt~. \label{erfcdef}
\end{eqnarray}
A plot of the universal function $H_S$, that is the symmetric part
of the wake in units of $\gamma/\sigma_\bot$, as a function of
$\Delta \xi$ is given in Fig. \ref{gs}.

Note that Eq. (\ref{sym4}) can be verified in the asymptotic limit
for $(\gamma \Delta s)^2 \gg \sigma_\bot^2$, i.e. $(\Delta \xi)^2
\gg 1$. If we further assume $(\Delta s)^2 \gg \sigma_z^2$ we
obtain the asymptote for two particles separated by a distance
$\Delta s$. It may be useful to remind that in agreement with
\cite{WAKK}, the case of a test particle in front of the source
corresponds to negative values of $\Delta s$. In this case, the
longitudinal electron density distribution and its Fourier
transform may be written respectively as

\begin{eqnarray}
f(t) = (-e)  \left[\delta\left(t-\frac{\Delta s}{2\beta
c}\right)+\delta\left(t+\frac{\Delta s}{2\beta c}\right)\right]
\label{2pdist}
\end{eqnarray}
and

\begin{eqnarray}
f(\omega) = (-e)  \left\{\exp\left[i \omega \frac{\Delta s}{2\beta
c}\right]+\exp\left[-i \omega \frac{\Delta s}{2\beta
c}\right]\right\} ~.\label{2pdis}
\end{eqnarray}
Substitution of Eq. (\ref{2pdis}) and Eq. (\ref{rhotrasf}) into
Eq. (\ref{repoyright3}) and use of Eq. (\ref{elost0}) yields

\begin{eqnarray}
\Delta \mathcal{E}_\mathrm{tot} = \frac{2  e^2}{\pi c}
\int_{-\infty}^{\infty} d \omega \left\{1+\cos \left[\frac{\omega
(\Delta s)}{ c}\right]\right\} \int_0^{\infty} d \theta
 \exp\left[-\frac{\theta^2 \omega^2
\sigma_\bot^2}{c^2}\right] \frac{\gamma^4
\theta^3}{(1+\gamma^2\theta^2)^2}.\cr \label{detotck}
\end{eqnarray}
The first term in parenthesis $\{...\}$ is related to
single-particle radiation in the far zone, while the second term
in $\cos(\cdot)$ may be interpreted as the interference term
between the field radiated by the two particles in the far zone.
Let us consider the interference term alone, that will be
indicated with $\Delta \mathcal{E}_\mathrm{int}$. Performing first
the integral in $d \omega$ we obtain

\begin{eqnarray}
\Delta \mathcal{E}_\mathrm{int} = \frac{2 e^2}{ \sqrt{\pi}
\sigma_\bot } \int_0^{\infty} d \theta \exp\left[-\frac{(\Delta
s)^2}{4 \theta^2 \sigma_\bot^2}\right]\frac{\gamma^4
\theta^2}{(1+\gamma^2\theta^2)^2}~.\label{detotck2}
\end{eqnarray}
In the limit for $(\gamma \Delta s)^2 \gg \sigma_\bot^2$, Eq.
(\ref{detotck2}) can be written as

\begin{eqnarray}
\Delta \mathcal{E}_\mathrm{int} = \frac{2 e^2}{ \sqrt{\pi}
\sigma_\bot } \int_0^{\infty} d x \exp\left[-\frac{x^2 (\Delta
s)^2}{4  \sigma_\bot^2}\right] = \frac{2 e^2}{\Delta s}
~,\label{detotck3}
\end{eqnarray}
where we have performed a change of integration variable to
$x=1/\theta$. The energy term $\Delta \mathcal{E}_\mathrm{int}$ is
due to the interaction between the two particles, and corresponds
to the energy dissipated by the system through the active part of
the wake.

Now, from Eq. (\ref{sym3}) we have that the energy lost by one
particle is $e^2 G_S(\Delta s) \longrightarrow -e^2/(\Delta s)$
when $(\gamma \Delta s)^2 \gg \sigma_\bot^2$. Since the system is
composed by two electrons and the wake $G_s$ is symmetric, the
total energy lost by the system is given by twice this value.
Thus, in the limit $(\gamma \Delta s)^2 \gg \sigma_\bot^2$, the
energy dissipated by the system due to the active part of the wake
is

\begin{eqnarray}
2 e^2 G_S(\Delta s) = - \frac{2 e^2}{\Delta s} ~,\label{detotck4}
\end{eqnarray}
in agreement with Eq. (\ref{detotck3}), as it must be.

The energy gained or lost by a single particle at position $s$
within the bunch due to the active (symmetric) part of the wake,
averaged over transverse coordinates is given by

\begin{eqnarray}
\Delta \mathcal{E}_S(s) = (-e) \int_{-\infty}^{\infty} G_S(\Delta
s) f(s-\Delta s) d (\Delta s)~. \label{elosss1}
\end{eqnarray}
Note that this expression can also be presented in terms of
impedances as

\begin{eqnarray}
\Delta \mathcal{E}_S(s) = \frac{1}{2\pi} \int_{-\infty}^{\infty}
d\omega~ \mathrm{Re}[Z](\omega) \bar{f}(\omega) \exp\left[-i
\omega \left(\frac{\Delta s}{\beta c}\right)\right]~,
\label{symens}
\end{eqnarray}
which however appears not to yield any computational advantage
over Eq. (\ref{elosss1}). Analogous remark holds for the
antisymmetric part of the wake.

With the help of Eq. (\ref{fs}) and Eq. (\ref{sym4}), we can
explicitly write $\Delta \mathcal{E}_S/\mathcal{E}_o$,
$\mathcal{E}_o = \gamma m_e c^2$ being the nominal energy of an
electron (with rest mass $m_e$) as a function of $\xi = \gamma
s/\sigma_\bot$:

\begin{eqnarray}
\frac{\Delta \mathcal{E}_S}{\mathcal{E}_o}(\xi) &=&
\frac{I_\mathrm{max}}{ \gamma I_A} \int_{-\infty}^{\infty} d
(\Delta \xi) H_S(\xi-\Delta \xi) \exp\left[-\frac{(\Delta
\xi)^2}{2 \eta^2}\right]~, \label{elosss2}
\end{eqnarray}
where we introduced the new parameter $\eta = \gamma
\sigma_z/\sigma_\bot $, and we called the Alfven current $I_A =
e/(m c^3) \simeq 17$ kA and the beam current $I_\mathrm{max} = e N
c/(\sqrt{~2\pi}~\sigma_z)$.

From a methodological point of view it is interesting to conclude
our treatment of the symmetric part of the wake with an additional
remark. On the one hand we have found that electrons in the bunch
lose energy under the action of radiative interaction forces
within an overtaking length $\sim 2 \gamma^2 \sigma_z$
\textit{after} the accelerator. On the other hand, an observer in
the far zone detecting the radiation pulse may calculate the
retarded position of radiators, concluding that electrons radiated
\textit{inside} the accelerator. Methodological questions of this
kind (electrons radiating when the radiative force does not work
on them and vice versa) are related with other well-known
situations of interest. For example, they also arise in Coherent
Synchrotron Radiation (CSR) from bending magnets and were
discussed in \cite{SALS}.

\subsection{\label{sub:antisy} Antisymmetric part of the wake}

\begin{figure}
\begin{center}
\includegraphics*[width=140mm]{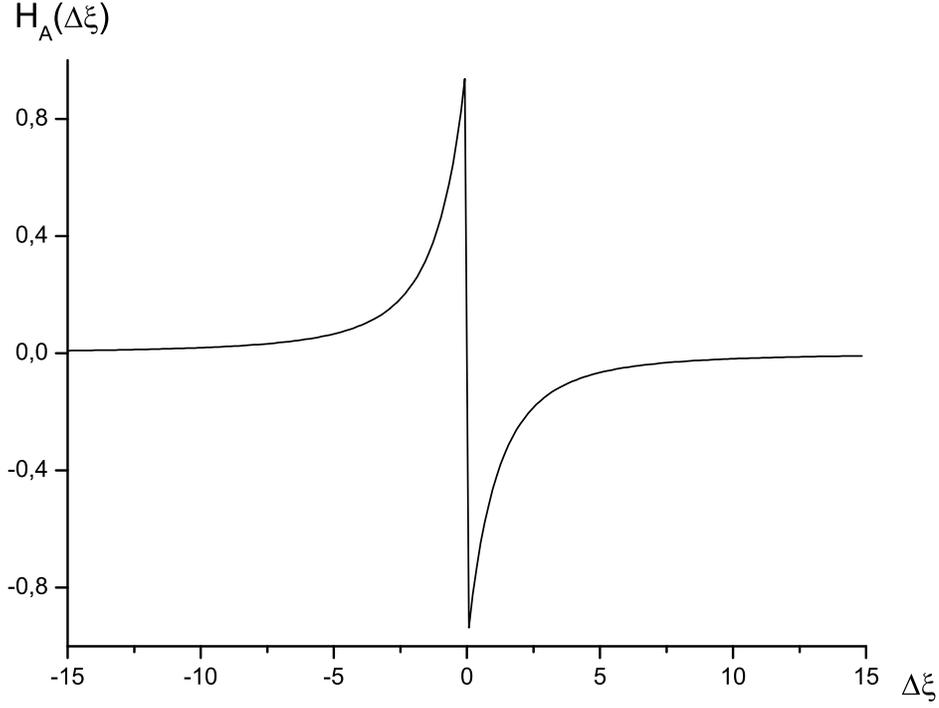}
\caption{\label{ga} Plot of the universal function $H_A$ as a
function of $\Delta \xi$. }
\end{center}
\end{figure}

The antisymmetric part of the wake function $G_S$ can be found
calculating

\begin{eqnarray}
G_A(\Delta s) = \frac{i}{2\pi} \int_{-\infty}^{\infty} d\omega~
\mathrm{Im}[Z](\omega) \exp\left[-i \omega \left(\frac{\Delta
s}{\beta c}\right)\right]~. \label{asym1}
\end{eqnarray}
With the help of Eq. (\ref{Zim}) and Eq. (\ref{rhotrasf}) one can
rewrite Eq. (\ref{asym1}) as

\begin{eqnarray}
G_A(\Delta s) = -  \frac{i z}{\pi c^2} \int_0^{\infty} \frac{d
{\theta} ~\theta} {1+\gamma^2\theta^2} \int_{-\infty}^{\infty}
d\omega~ \omega\exp\left[-\frac{\theta^2 \omega^2
\sigma_\bot^2}{c^2}\right] \exp\left[-i \omega \left(\frac{\Delta
s}{c}\right)\right]~. \label{asym2}
\end{eqnarray}
Using the fact that

\begin{eqnarray}
&&\int_{-\infty}^{\infty} d\omega~ \omega\exp\left[-\frac{\theta^2
\omega^2 \sigma_\bot^2}{c^2}\right] \exp\left[-i \omega
\left(\frac{\Delta s}{c}\right)\right] = \cr && i c \frac{d}{d
(\Delta s)} \int_{-\infty}^{\infty} d\omega~
\exp\left[-\frac{\theta^2 \omega^2 \sigma_\bot^2}{c^2}-i \omega
\left(\frac{\Delta s}{c}\right)\right]= - \frac{i \sqrt{\pi} ~c^2
\Delta s}{2 \theta^3 \sigma_\bot^3} \exp\left[-\frac{(\Delta
s)^2}{4 \theta^2 \sigma_\bot^2}\right]~, \cr &&\label{relasy}
\end{eqnarray}
one obtains

\begin{eqnarray}
G_A(\Delta s) =   - \frac{~z ~ \Delta s}{ 2 \sqrt{\pi}
 \sigma_\bot^3}\int_0^{\infty} \frac{d {\theta} }
{(1+\gamma^2\theta^2)\theta^2} \exp\left[-\frac{(\Delta s)^2}{4
\theta^2 \sigma_\bot^2}\right]~. \label{asym3}
\end{eqnarray}
Performing the integral in $d\theta$ and using, as before,  the
notation $\Delta \xi = \gamma (\Delta s)/\sigma_\bot$ one has

\begin{eqnarray}
G_A(\Delta \xi) =   \frac{\gamma \eta \hat{z}}{\sigma_\bot}
~H_A(\Delta \xi) \label{asym4}
\end{eqnarray}
where

\begin{eqnarray}
H_A(\Delta \xi) =   - \frac{1}{2 \sqrt{\pi}} (\Delta
\xi)~\left\{2\frac{\sqrt{\pi}}{|\Delta \xi|} - \pi \exp
\left[\frac{ (\Delta\xi)^2}{4}\right]
\mathrm{erfc}\left[\frac{|\Delta\xi|}{2}\right]\right\}~,
\label{asym4b}
\end{eqnarray}
and we redefined $\hat{z} = {z}/(2\gamma^2 \sigma_z)$. A plot of
the universal function $H_A$, that is the symmetric part of the
wake in units of $\gamma \hat{z}/\sigma_\bot$, as a function of
$\Delta \xi$ is given in Fig. \ref{ga}.

Note that Eq. (\ref{asym4}) can be verified in the asymptotic
limit $(\gamma \Delta s)^2 \gg \sigma_\bot^2$, i.e. $(\Delta
\xi)^2 \gg 1$. If we further assume $(\Delta s)^2 \gg \sigma_z^2$
we obtain the asymptote for two particles separated by a distance
$\Delta s$ as before for the symmetric part of the wake. $\Delta
s<0$ indicates again the case when the test particle is in front
of the source. From the Lienard-Wiechert expression for the
electric field of a single particle moving on a straight line we
can calculate the rate of energy change of the front electron

\begin{equation}
\frac{d\mathcal{E}}{dz} = \frac{e^2}{\gamma^2 (\Delta
s)^2}~.\label{checkG1}
\end{equation}
Independently, from Eq. (\ref{asym4}) we have that $H_A(\Delta
\xi) \longrightarrow  - 2 ~\mathrm{sign}(\Delta \xi)/(\Delta
\xi)^2$ when $(\Delta \xi)^2 \gg 1$. As a result, at position $z$
the front particle has gained the energy

\begin{equation}
\Delta \mathcal{E}_A = e^2 G_A(\Delta s) = \frac{ e^2 z}{ \gamma^2
(\Delta s)^2 }\label{checkG}
\end{equation}
that corresponds to the rate of energy change in Eq.
(\ref{checkG1}) as it must be.

The energy gained or lost by a single particle at position $s$
within the bunch due to the reactive part of the wake, averaged
over transverse coordinates is given by

\begin{eqnarray}
\Delta \mathcal{E}_A(s) = (-e) \int_{-\infty}^{\infty} G_A(\Delta
s) f(s-\Delta s) d (\Delta s)~. \label{elossa1}
\end{eqnarray}
With the help of Eq. (\ref{fs}) and Eq. (\ref{asym4}), we can
explicitly write $\Delta \mathcal{E}_A/\mathcal{E}_o$,
$\mathcal{E}_o = \gamma m_e c^2$ being the nominal energy of an
electron (with rest mass $m_e$) as a function of $\xi = \gamma
s/\sigma_\bot$:

\begin{eqnarray}
\frac{\Delta \mathcal{E}_A}{\mathcal{E}_o}(\xi) &=&
\frac{I_\mathrm{max}}{ \gamma I_A} ~\eta \hat{z}
\int_{-\infty}^{\infty} d (\Delta \xi) H_A(\xi-\Delta\xi)
\exp\left[-\frac{(\Delta \xi)^2}{2 \eta^2}\right]~.\label{elossa2}
\end{eqnarray}
Note that Eq. (\ref{elossa2}) is a function of $\xi$ but also
depends parametrically on $\eta$, and may be presented as

\begin{eqnarray}
\frac{\Delta
\mathcal{E}_A}{\mathcal{E}_o}\left(\frac{s}{\sigma_z}; \eta\right)
&=& \frac{I_\mathrm{max} \hat{z}}{ \gamma I_A}
F\left(\frac{s}{\sigma_z}; \eta\right)~.\label{elossa3}
\end{eqnarray}
where we indicated the parametric dependence of $\eta$ after the
semicolon and

\begin{figure}
\begin{center}
\includegraphics*[width=140mm]{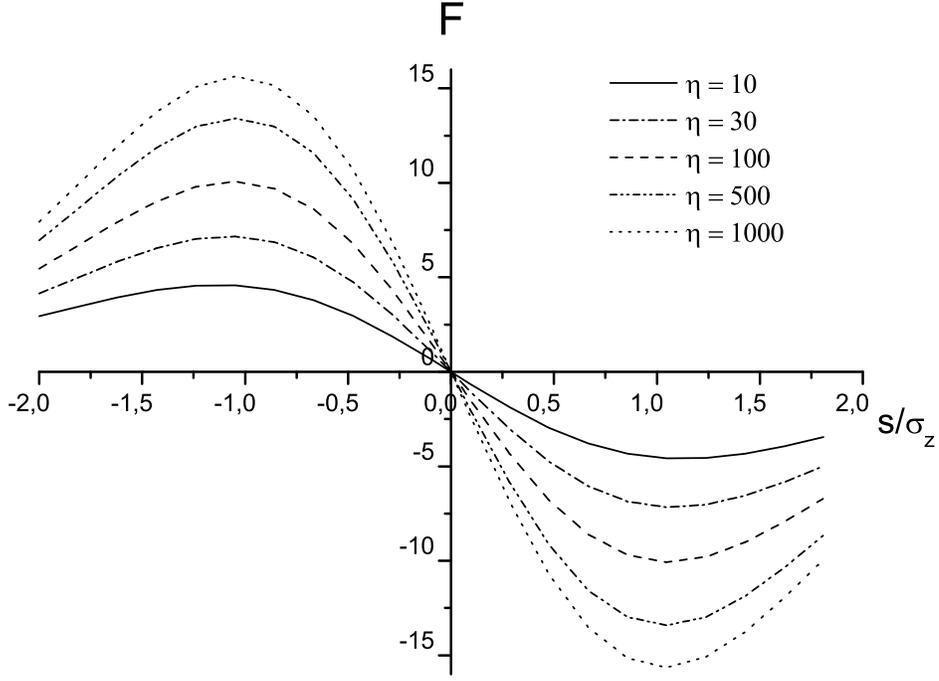}
\caption{\label{fseta} Plot of F in Eq. (\ref{elossaker}) as a
function of $s/\sigma_z$ for different values of $\eta$.}
\end{center}
\end{figure}

\begin{eqnarray}
F\left(\frac{s}{\sigma_z}; \eta\right) &=& \int_{-\infty}^{\infty}
d (\Delta \xi)~ \eta~ H_A\left(\eta
\frac{s}{\sigma_z}-\Delta\xi\right) \exp\left[-\frac{(\Delta
\xi)^2}{2 \eta^2}\right]~.\label{elossaker}
\end{eqnarray}
A plot of Eq. (\ref{elossaker}) is given as a function of
$s/\sigma_z$ in Fig. \ref{fseta} for different values of $\eta$.

We failed to integrate Eq. (\ref{elosss2}) and Eq. (\ref{elossa2})
(or Eq. (\ref{elossa3})) analytically. However, plots for ${\Delta
\mathcal{E}_S}/{\mathcal{E}_o}$ and ${\Delta
\mathcal{E}_A}/{\mathcal{E}_o}$ as a function of $\xi$ can be
computed with the help of numerical techniques. In the following
Section we will give a practical example of application of our
work.

\section{\label{sec:example} Impact on the
design of a table-top FEL in the VUV and X-ray range}

The foreseen specifications of the next generation laser-plasma
accelerators are stunning.  The high acceleration gradient will be
up to TV/m, producing very short bunches about $10$ fs long in the
$100$ kA current class (i.e. with a charge of about $1$ nC). The
beam quality is also expected to rival state-of-the art
conventional acceleration techniques, featuring a relative energy
spread of $0.1\%$ for beam energies in the GeV range and a
normalized emittance in the order of $1$ mm$\cdot$mrad.

One of the obvious and perhaps most exciting applications
envisaged for these machines is as drivers for table-top FELs,
both in the VUV (TT-VUV FEL) and in the X-ray range (TT-XFEL)
\cite{TTBU}. Estimation of basic scaling parameters for
Self-Amplified Spontaneous Emission (SASE) FEL applications within
the one-dimensional ideal case indicate that a system composed by
a laser-plasma accelerator and a meter-long undulator may undergo
SASE process in the sub-nanometer range, thus outclassing all
existing and foreseen XFEL projects, both in dimensions and costs.

Unfortunately, the road map towards the practical realization of
this ambitious goal is not problem-free. Let aside other possible
problems we focus our attention on a fundamental issue that, in
our understanding, constitutes a serious difficulty. Namely, we
want to estimate the impact of the longitudinal wake on the
electron beam quality using the theory developed in the previous
Sections. This is possible, because the plasma acceleration
gradient is such that the electron beam to be injected in the
undulator can be produced within distances $d_a \ll 2 \gamma^2
\sigma_z$, i.e. well within the applicability region of our
theory. Note that our estimation is completely independent of the
physical realization of the accelerator (a laser-plasma device, or
any other physical principle). Within our model, the accelerator
is just located at a single point at position $z_A$ down the
beamline.

\begin{figure}
\begin{center}
\includegraphics*[width=140mm]{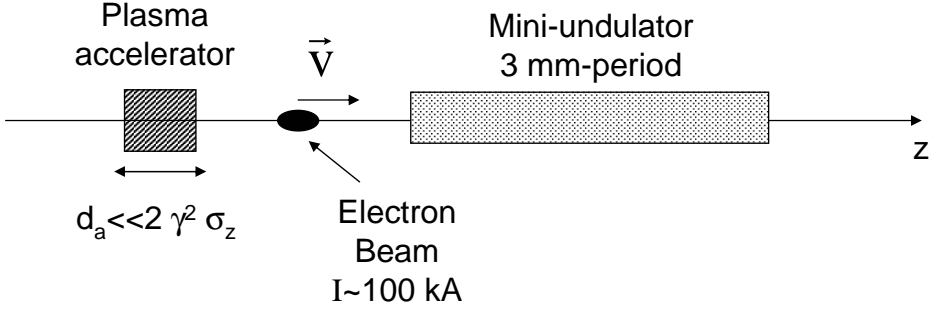}
\caption{\label{geom3} Geometry of the TT-VUV FEL and of the
TT-XFEL setups.}
\end{center}
\end{figure}

In the following we will consider two sets of parameters
informally under discussion within the scientific community. The
first refers to a test case designed to radiate in the VUV range
($\lambda = 25$ nm) while the second deals with the true table-top
SASE XFEL, designed to radiate in the hard X-ray range ($\lambda =
0.25$ nm). Both parameter sets rely on the use of a mini-undulator
with a $3$ mm-period, that would allow low electron beam energy
(in the order $100$ MeV and $1$ GeV for the VUV and the X-ray
case). Very high currents, in the order of $100$ kA, are supposed
to be used . The geometry of the system is described in Fig.
\ref{geom3}. It is important to note that in both study-cases the
undulator parameter $K \simeq 0.5$. This means that the
longitudinal velocity of particles in the undulator is not
significatively altered ($\gamma^2_z \equiv \gamma^2/(1+K^2/2)
\simeq 1.125 \gamma^2$). As a result, wake field calculations in
vacuum, performed in the preceding Sections, can still be used
inside the undulator.

Description of the FEL setups will be limited to the
one-dimensional model. In this case, the basic scaling parameter
is the one-dimensional $\rho_\mathrm{1D}$ parameter defined as
\cite{PELL}

\begin{eqnarray}
\rho_\mathrm{1D} = \frac{\lambda_w}{4\pi} \left[\frac{2 \pi^2 j_o
K^2 A^2_{JJ} }{I_A \lambda_w \gamma^3}\right]^{1/3} \label{rhopar}
\end{eqnarray}
where $j_o$ is the beam current density, $\lambda_w$ is the
undulator period, $K$ is the undulator parameter defined as $K= e
\lambda_w H_w/(2\pi m_e c^2)$, where $H_w$ is the maximum magnetic
field produced by the undulator on the $z$ axis. Finally, the
coupling factor $A_{JJ}$, for a planar undulator, is given by
$A_{JJ} = J_0(Q)-J_1(Q)$, where $Q = K^2/(2+K^2)$. The main
quantities of interest characterizing the FEL process can be
written in terms of $\rho_\mathrm{1D}$. In fact, the
one-dimensional power gain length of a mono-energetic beam is

\begin{eqnarray}
L_\mathrm{G} = \frac{\lambda_w}{4\pi \sqrt{3} \rho_\mathrm{1D}}~,
\label{LG}
\end{eqnarray}
while the relative FEL bandwidth at saturation is close to
$\rho_\mathrm{1D}$ and the power at saturation is about
$\rho_\mathrm{1D}$ times the electron beam power.

Let us first deal with the VUV test case, that aims at producing
radiation at a wavelength $\lambda = 25$ nm using an electron
energy $\mathcal{E}_o = 130$ MeV and a $0.7$ m-long undulator with
period $\lambda_w = 3$ mm. The electron beam current is about $60$
kA. The longitudinal beam size is taken to be $\sigma_z \sim 1
~\mu$m. The transverse beam dimension is changed, instead, from an
initial size $\sigma_\bot = 1 ~\mu$m to a final size $\sigma_\bot
\sim 30 ~\mu$m within a few centimeters. In our estimation we will
neglect the impact of this change of dimension on the longitudinal
wake, and take $\sigma_\bot \sim 30 ~\mu$m from the very
beginning.

\begin{figure}
\begin{center}
\includegraphics*[width=140mm]{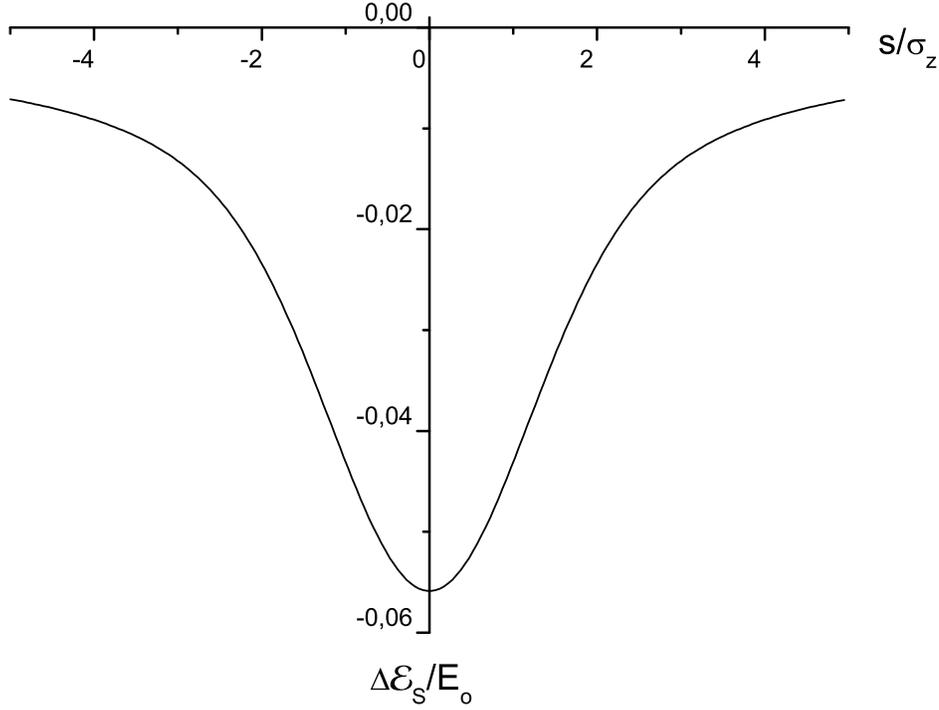}
\caption{\label{tests} Relative energy change $\Delta
\mathcal{E}_s/\mathcal{E}_o$ as a function of the position inside
the bunch $s/\sigma_s$ in the case of the VUV table top FEL
setup.}
\end{center}
\end{figure}
\begin{figure}
\begin{center}
\includegraphics*[width=140mm]{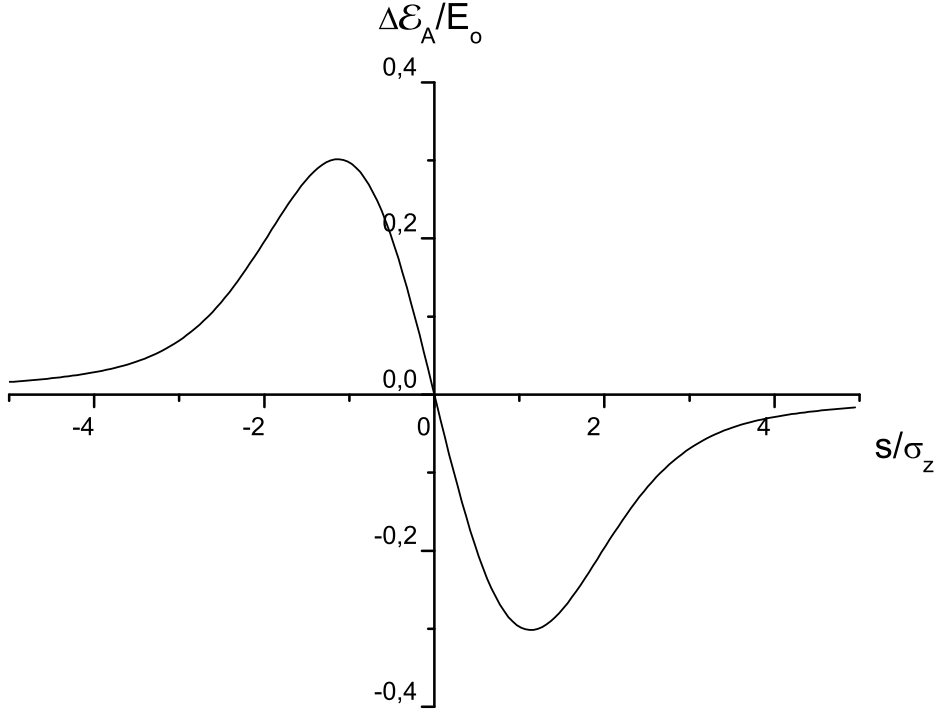}
\caption{\label{testa} Relative energy change $\Delta
\mathcal{E}_A/\mathcal{E}_o$ as a function of the position inside
the bunch $s/\sigma_s$ in the case of the VUV table top FEL setup.
This plot refers to the position $z=0.7$ m. }
\end{center}
\end{figure}
The radiation formation length at wavelength $\lambda$ is
estimated to be $2 \gamma^2 \lambdabar$, and our asymptotic
calculations for the impedance are valid for $z \gg 2 \gamma^2
\lambdabar$. An estimation of the wavelengths of interest is given
considering the typical length scale that enters in the
expressions for the impedance. In our case we take $\lambdabar
\simeq \sigma_z = 1 ~\mu$m, which yields $2 \gamma^2 \lambdabar
\simeq 10$ cm, a few times smaller than the size of the TT-VUV
FEL, because the planned length of the undulator is about $0.7$ m.
This allows us to use asymptotic expressions Eq. (\ref{elosss2})
and Eq. (\ref{elossa2}). The only parameters needed to be plugged
into these equations are thus $\gamma = 260$ and $\eta = 8.7$ and
$I = 60$ kA. Moreover, since the asymmetric part of the wake is
proportional to $\hat{z}=z/(2 \gamma^2 \sigma_z)$, we need to fix
a position along the longitudinal axis to calculate Eq.
(\ref{elossa2}). Here we set $z \simeq 0.7$ m, corresponding to
the foreseen undulator length.  The relative energy change $\Delta
\mathcal{E}_{S,A}/\mathcal{E}_o$ calculated with the help of Eq.
(\ref{elosss2}) and Eq. (\ref{elossa2}) is plot respectively in
Fig. \ref{tests} and Fig. \ref{testa} as a function of the
longitudinal coordinate inside the bunch. While travelling down
the undulator, a correlated energy change develops along the
electron beam. In our case the chirp is non-linear, but in order
to estimate the magnitude of the effect we can use the linear
energy chirp parameter \cite{KRIS,SALC}:

\begin{equation}
\hat{\alpha} = -\frac{d \gamma}{dt} \frac{1}{\gamma \omega
\rho_{1D}^2} ~.\label{alphas}
\end{equation}
The effect of linear energy chirp starts to play a significant
role on the FEL gain when $\hat{\alpha} \gtrsim 1$. Intuitively,
this means that the relative energy change  becomes comparable
with the FEL parameter on the scale of the coherence length. Using
an estimation of the slope around $s = 0$ in Fig. \ref{testa} and
the other problem parameters one finds $\hat{\alpha} \simeq 12$,
that indicates a very large effect. It is important to note that
in our case the energy chirp in Fig. \ref{testa} depends on the
electron beam profile but also on time, because it develops along
the undulator. This effect is fundamental, it cannot be avoided,
and is directly linked with the feasibility of the proposed FEL
scheme. In other words, radiation generated in one part of the
undulator cannot interact in resonance in another part of the
undulator, and the amplification process is destroyed.

In these conditions, our conclusion is that the proposed TT-VUV
FEL setup will not work.

Let us now turn to consider a table-top SASE XFEL scenario aimed
at producing radiation at a wavelength $\lambda = 0.25$ nm using
an electron energy $\mathcal{E}_o = 1.2$ GeV and a $3$m-long
undulator with period $\lambda_w = 3$ mm. The electron beam
current is taken in the $100$ kA range. We take again $\sigma_z =
1 ~\mu$m, and $\sigma_\bot = 30 ~\mu$m. Similarly as before we
estimate the radiation formation length taking $\lambdabar \simeq
\sigma_z$. This yields $2 \gamma^2 \lambdabar \simeq 10$ m, longer
than the TT-XFEL undulator length, that is about $3$ m. This makes
it impossible to use our asymptotic expressions, Eq.
(\ref{elosss2}) and Eq. (\ref{elossa2}), because the entire setup
is well within the formation length.  This situation needs further
study, based on detailed numerical simulations, that goes beyond
the scope of this work. However, as before, we can say that
longitudinal wake fields will be responsible for an energy chirp
that is  both profile dependent and time dependent. Moreover, even
though the electron beam energy has increased of an order of
magnitude (thus leading to a decrease of the energy change level),
the $\rho_\mathrm{1D}$ parameter is decreased of an order of
magnitude. Since the undulator length is only a factor three
shorter than the formation length, we conclude that wake fields
constitute a major effect in this case too. Such effect poses a
serious threat to the operation of the TT-XFEL setup.

\section{\label{sec:conc} Conclusions}

This work constitutes the first study of the impact of
longitudinal wake fields on the quality of electron beams produced
with high field-gradient accelerators. We restricted our attention
to the analysis of the wake generated along the trajectory
following the acceleration, and assuming that the acceleration
happens on a short longitudinal scale compared with the overtaking
length, i.e. $d_a \ll 2 \gamma^2 \sigma_z$. Thus, our
consideration does not depend on the particular realization of the
accelerator. However, given present technological developments,
one of the most relevant applications of our study should come
from the realm of laser-plasma accelerators.

We calculated longitudinal symmetric and anti-symmetric parts of
the wake function as well as real and imaginary part of the
impedance with the help of paraxial approximation within a
space-frequency domain formulation of Maxwell's equation. While
the general expressions for wakes and impedances need numerical
techniques to be evaluated, it is possible to present analytical
expressions for the asymptotic limit when the electron bunch has
reached a position, down the beamline, that is far from the
acceleration point, with respect to the overtaking distance. It
should be noted that in these cases the wake and the impedance are
proportional to universal functions, valid for any set of
parameters.

Our results can be used as analytical benchmarks for computer
codes. Moreover, the the asymptotic result for the antisymmetric
part of the wake is of fundamental importance, in the sense that
its validity is not restricted to the particular model studied in
this paper (fast acceleration). To the best of our knowledge, Eq.
(\ref{asym4}) constitutes the first analytical solution to the
problem of space-charge wakes (calculated per unit length). Taking
advantage of similarity techniques we presented such result in
terms of a dimensionless object dependent on the problem
parameters $(\gamma \eta \hat{z}/\sigma_\bot)$ multiplied by a
universal function $H_A$ (see Eq. (\ref{asym4b})).  Eq.
(\ref{asym4}) can further be used to calculate the relative energy
change of a particle within a given bunch. In case of a Gaussian
longitudinal profile the relative energy spread reduces to Eq.
(\ref{elossa3}). Plots presented in Fig. \ref{fseta} will help the
reader to perform a fast estimation of the influence of the
space-charge wake at the stage of planning of experiments.

As a particular example, we used our results to estimate the
impact of the longitudinal wake fields on the energy change of the
electron beam in a table top FEL setup, both in the VUV range
($\lambda= 25$ nm) and in the x-ray range ($\lambda = 0.25$ nm).
In both situations, such energy variation is time and shape
dependent. This effect is fundamental, in the sense that it cannot
be avoided. The total energy deviation is found to be an important
effect, and constitute a major problem for the exploitation of the
SASE amplification process.

\section{\label{sec:graz} Acknowledgements}

We thank Massimo Altarelli and Martin Dohlus (DESY) for many
useful discussions, Reinhard Brinkmann and Jochen Schneider (DESY)
for their interest in this work.

\end{document}